# VIRTUAL SCREENING AND LEAD OPTIMIZATION TO IDENTIFY NOVEL INHIBITORS FOR HDAC-8

**Maria Antony Dhivyan JE and Anoop MN**

School of Health and Life Sciences, Edinburgh Napier University, Edinburgh, United Kingdom – EH10 5DT

## ABSTRACT

Histone deacetylase (HDAC) and Histone acetyl-transferase (HAT) are enzymes that influence transcription by selectively deacetylating or acetylating the ε-amino groups of lysine located near the amino termini of core histone proteins. Over expression of HDACs noted in many forms of cancers including leukemia and breast cancer. HDAC inhibitors have been shown to be potent inducers of growth arrest, differentiation, and/or apoptotic cell death. There is a growing interest in the development of histone deacetylase inhibitors as anti cancer agents. Three known ligands of HDAC-8 were taken and docked. The best scores were analyzed and structures similar to these ligands were downloaded using c@rol and corina databases and docked. Also large databases of small molecules were computationally screened using molecular docking for "hits" that can conformationally and chemically fit to the active site. Molecules which got high scores for both GoldScore and ChemScore were selected and compared with the previous results. Those with best results were then taken for calculating H-bond interactions and close contacts. Bioactivity prediction of the best ranked ligands was done. Their physicochemical properties were also analyzed. Four new molecules were identified and suggested for further testing in the wet lab.

# INTRODUCTION

Bioinformatics involve the use of techniques including applied mathematics, informatics, statistics, computer science, artificial intelligence, chemistry, and biochemistry to solve biological problems usually on the molecular level. Major research efforts in the field include sequence alignment, gene finding, genome assembly, protein structure alignment, protein structure prediction, prediction of gene expression and protein-protein interactions, and the modeling of evolution.

Bioinformatics can influence significantly in solving the following types of problems:

1. Prediction of 3-D structure based on linear genomic information, i.e., the study of structural genomics.
2. Gene expression analysis, prediction of gene function and establishment of gene libraries (functional genomics).
3. The ability to use genome sequence to identify proteins and their functions, protein interactions, modifications and functions, i.e., the field of proteomics.
4. Simulating metabolism from the biochemical functions of an organism.
5. Molecular modeling and molecular dynamics are the methods to predict structure from experimental data.
6. Data obtained from functional genomics and proteomics could be used in drug designing and discovery.

With the near completion of the human genome sequencing, bioinformatics has established itself as an essential tool in target discovery and the insilico analysis of gene expression and gene function are now an integral part of it, facilitating the selection of the most relevant targets for a disease under study.

A bulk of techniques, both old and new, has recently matured into potent weapons in the war against disease. The need for ongoing development of new drugs needs no emphasis in the light of the current global situation of health and disease

Drug discovery research relies heavily on bioinformatics to manage the databases of small molecules that are potential lead compounds, to search databases of protein structures for structure-based drug design methods, and to model the docking of compounds and their target proteins.

The need for ongoing development of new drugs needs to emphasis in light of the current global situation of health and disease. Traditionally, the process of drug development has revolved around a screening approach, as nobody knows which compound or approach could serve as a drug or therapy. Such almost blind screening approach is very time-consuming and laborious. The short coming of traditional drug discovery; as well as the allure of a more deterministic approach to combating disease has led to the concept of "Rational drug design"(Kuntz 1992).

Nobody could design a drug before knowing more about the disease or infectious process than past. For "rational" design, the first necessary step is the identification of a molecular target critical to a disease process or an infectious pathogen. Then the important prerequisite of "drug design" is the determination of the molecular structure of target, which makes sense of the word "rational". In fact, the validity of "rational" or "structure-based" drug discovery rests largely on a high-resolution target structure of sufficient molecule detail to allow selectivity in the screening of compounds.

## 1.1 COMPUTER-AIDED DRUG DESIGN (CADD)

Computer-aided drug design (CADD), is also called computer-aided molecular design (CAMD), represents more recent applications of computer as tools in the drug design process. This field includes computer graphics, 3-D model of molecules (Molecular Modeling), protein structure prediction and analysis, molecular motion (Molecular dynamic simulation), molecular shape(conformational analysis), molecular property prediction, quantitative structure/property relationships (QSAR/QSPR), database search, quantum chemistry(for predicting structure properties and reactivity), computer assisted synthesis, protein/drug "docking" etc. the techniques provided by computational methods include computer graphics for visualization and the methodology of theoretical chemistry.

The development of new drugs with potential therapeutic applications is one of the most complex and difficult process in the pharmaceutical industry. Millions of dollars and man-hours are devoted to the discovery of new therapeutic agents. As, the activity of a drug is the result of a multitude of factors; rational drug design has been utopias for centuries. Computers have magnified the capability of scientists to collect, access, and analyze information, and even do "virtual" experiments. Recently Computer-based drug design (CADD) has caused a "quiet explosion" in modern drug discovery. CADD can guide and assist the design of new therapeutic agents with desired properties by means of molecular modeling, theoretical calculation and prediction methods. The aim of using the computer for drug design is to analyze the interactions between the drug and its receptor site and to "design" molecules that give an optimal fit. The central assumption is that a good fit results from structural and chemical complementarities to the target receptor.

The steps involved are:

### 1.1.1 *TARGET IDENTIFICATION*

This step aims to identify a biological drug target. This is typically a receptor, enzyme or ion channel that needs to be manipulated to prevent the development of a disease or alleviate symptoms. Drug usually act on either cellular or genetic chemicals in the body, known as targets, which are believed to be associated with disease. Scientists use a variety of techniques to identify and isolate a target and learn more about its functions and how these influence disease. Compounds are then identified that have various interactions with drug targets helpful in treatment of a specific disease. Thus, we concentrate our efforts on discovering or even inventing compounds that can alter the disease-causing mechanism, whether a single protein or a complex pathway of proteins, to bring it back into line with normal function.

## 1.1.2 TARGET VALIDATION

To select targets most likely to be useful in the development of new treatments for disease, researchers analyze and compare each drug target to others based on their association with a specific disease and their ability to regulate biological and chemical compounds in the body. Tests are conducted to confirm that interactions with the drug target are associated with a desired change in the behavior of diseased cells. Research scientists can then identify compounds that have an effect on the target selected.

## 1.1.3 LEAD IDENTIFICATION

A lead compound or substance is one that believed to have potential to treat disease. Laboratory scientists can compare known substance with new compounds to determine their likelihood of success. Leads are sometimes developed as collections, or libraries, of individual molecules that possess properties needed in a new drug. The most important source of leads is "libraries" of molecule (e.g.) natural product libraries, peptide libraries, carbohydrates libraries, etc. "Virtual libraries" can be created by using combinatorial chemistry. Testing is then done on each of these molecules to confirm its effect on the drug target.
Some of the technologies used in the lead identification are:

1. Virtual screening
2. High throughput docking

### 1.1.3.1 VIRTUAL SCREENING

The dominant technique for the identification of new lead compounds in drug discovery is the physical screening of large libraries of chemicals against a biological target (high throughput screening). Virtual screening is an alternative approach is to computationally screen large libraries of chemicals for compounds that complement targets of known structure, and experimentally test those that are predicted to bind well. It access a large number of possible new ligands which can be purchased and tested. Virtual screening, or insilico screening, is a new approach attracting increasing levels of interest in the pharmaceutical industry as a productive and cost-effective technology in the search for novel lead compounds. Although the principles involved-the computational analysis of chemical databases to identify compounds appropriate for a given biological receptor-have been pursued for several years in molecular modeling groups, the availability of inexpensive high-performance computing platforms has transformed the process so that increasingly complex and more accurate analyses can be performed on very detailed and relevant basis for prioritizing compounds for biological screening. Virtual screening offers a practical route to discovering new reagents and lead for pharmaceutical research.

### 1.1.3.2 HIGH THROUGHPUT DOCKING:

Docking is research technique for predicting whether one molecule will bind to another, usually a protein. Docking is a term used for computational schemes that attempt to find the best matching between two molecules: receptor and a ligand. If the geometry of the pair is complimentary and involves favorable biochemical interactions, the ligand will potentially bind the protein (receptor).

### 1.1.4 LEAD OPTIMIZATION:

Lead optimization compares the properties of various lead compounds and provides information to select the compound or compounds with the greatest potential to be developed into safe and effective medicines. The candidate drugs with better therapeutic profiles are accessed for quality, taking into account factors such as the ease of synthesis and formulation. After this they are registered as an investigational new drug and submitted for clinical drug.

### 1.1.5 TESTING OF THE ACTIVE COMPOUND (PRE-CLINICAL PHASE)

After optimizing the active compound, testing in the preclinical phase lab and animal testing is used to verify whether it is principally suited for use in the human body. To determine this, the researchers examine among other things how the compound is absorbed by the body, how it is excreted, and how it affects the organs.

In addition, they examine whether and in concentration it has a toxic or effect on the genetic makeup.

### *1.1.6 CLINICAL TRIALS*

If the active compound proves successful and also fulfils the legal requirements, it is then directly tested on human beings in three clinical phases:

Phase 1: Compatibility in health test subjects

Phase 2: Determination of optimal doses

Phase 3: Proof of effectiveness.

### *1.1.7 APPROVAL PROCESS*

If the medicine has made its way through all of the preceding phases, the process of getting approval from the authorities begins. The drug cannot be marketed until approval has been obtained.

Target identification

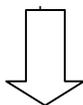

Target validation

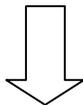

Lead identification

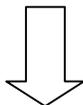

Lead optimization

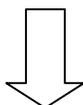

Clinical trials

## 1.2 DRUG

Drug is defined as "a chemical substance used in the treatment, cure, prevention, or diagnosis of disease or used to otherwise enhance physical or mental well-being. Drugs may be prescribed for a limited duration, or on a regular basis for chronic disorders.

## 1.3 TARGET

A drug target is a key molecule involved in a particular metabolic or signaling pathway that is specific to a disease condition or pathology, or to the infectivity or survival of a microbial pathogen. Drugs are used to stop the functioning of the pathway in the diseased state by causing a key molecule to stop functioning. Drugs may be designed that bind to the active region and inhibit this key molecule.

## 1.4 LIGAND

A ligand is a molecule that is able to bind and form a complex with a biomolecule to serve a biological purpose. It is an effector molecule binding to a site on a target protein, by intermolecular forces such as ionic bonds, hydrogen bonds and Van der Waal forces. The docking (association) is usually reversible (dissociation). Actual irreversible covalent binding between a ligand and its target molecule is rare in biological systems.. Ligand binding to receptors alters the chemical conformation, i.e. the three dimensional shape of the receptor protein. The conformational state of a receptor protein determines the functional state of a receptor. The tendency or strength of binding is called affinity. Ligands include substrates, inhibitors, activators, and neurotransmitters.

## 1.5 IC50

The **IC$_{50}$** is a measure of drug effectiveness. It indicates how much of a particular drug or other substance (inhibitor) is needed to inhibit a given biological process by half. In other words, it is the half maximal (50%) inhibitory concentration (IC) of a substance (50% IC, or IC$_{50}$). It is commonly used as a measure of antagonist drug potency in pharmacological research. IC$_{50}$ represents the concentration of a drug that is required for 50% inhibition *in vitro.*

## 1.6 DOCKING

Docking is the process by which two molecules fit together in 3D space. Docking is a method which predicts the preferred orientation of one molecule to a second when bound to each other to form a stable complex. Docking is frequently used to predict the binding orientation of small molecule drug candidates to their protein targets in order to in turn predict the affinity and activity of the small molecule. Hence docking plays an important role in the rational design of drugs. Molecular docking may be defined as an optimization problem, which would describe the "best-fit" orientation of a ligand that binds to a particular protein of interest.

The focus of molecular docking is to computationally stimulate the molecular recognition process. The aim of molecular docking is to achieve an optimized conformation for both the protein and ligand and relative orientation between protein and ligand such that the free energy of the overall system is minimized.

### *1.6.1 APPROACHES TO MOLECULAR DOCKING*

Two approaches are particularly popular within the molecular docking community. One approach uses a matching technique that describes the protein and the ligand as complementary surfaces. The second approach simulates the actual docking process in which the ligand-protein pair wise interaction energies are calculated. Both approaches have significant advantages as well as some limitations.

#### *1.6.1.1 SHAPE COMPLEMENTARITY METHODS*

Geometric matching/ shape compelementarity methods describe the protein and ligand as a set of features that make them dockable. These features may include molecular surface/ complementary surface descriptors. In this case, the receptor's molecular surface is described in terms of its solvent-accessible surface area and the ligand's molecular surface is described in terms of its matching surface description. The complementarity between the two surfaces amounts to the shape matching description that may help finding the complementary pose of docking the target and the ligand molecules. Another approach is to describe the hydrophobic features of the protein using turns in the main-chain atoms. Yet another approach is to use a Fourier shape descriptor technique described in [ref]. Whereas the shape complementarity based approaches are typically fast and robust, they cannot usually model the movements or dynamic changes in the ligand/ protein conformations accurately, although recent developments allow these methods to investigate ligand flexibility. Shape complementarity methods can quickly scan through several thousand ligands in a matter of seconds and actually figure out whether they can bind at the protein's active site, and are usually scalable to even protein-protein interactions. They are also much more amenable to pharmacophore based approaches, since they use geometric descriptions of the ligands to find optimal binding.

*1.6.1.2 SIMULATION PROCESSES*

The simulation of the docking process as such is a much more complicated process. In this approach, the protein and the ligand are separated by some physical distance, and the ligand finds its position into the protein's active site after a certain number of "moves" in its conformational space. The moves incorporate rigid body transformations such as translations and rotations, as well as internal changes to the ligand's structure including torsion angle rotations. Each of these moves in the conformation space of the ligand induces a total energetic cost of the system, and hence after every move the total energy of the system is calculated. The obvious advantage of the method is that it is more amenable to incorporating ligand flexibility into its modeling whereas shape complementarity techniques have to use some ingenious methods to incorporate flexibility in ligands. Another advantage is that the process is physically closer to what happens in reality, when the protein and ligand approach each other after molecular recognition. A clear disadvantage of this technique is that it takes longer time to evaluate the optimal pose of binding since they have to explore a rather large energy landscape. However grid-based techniques as well as fast optimization methods have significantly ameliorated these problems.

## 1.6.2 THE MECHANICS OF DOCKING

To perform a docking screen, the first requirement is a structure of the protein of interest. Usually the structure has been determined using a biophysical technique such as x-ray crystallography, or less often, NMR spectroscopy. This protein structure and a database of potential ligands serve as inputs to a docking program. The success of a docking program depends on two components: the search algorithm and the scoring function.

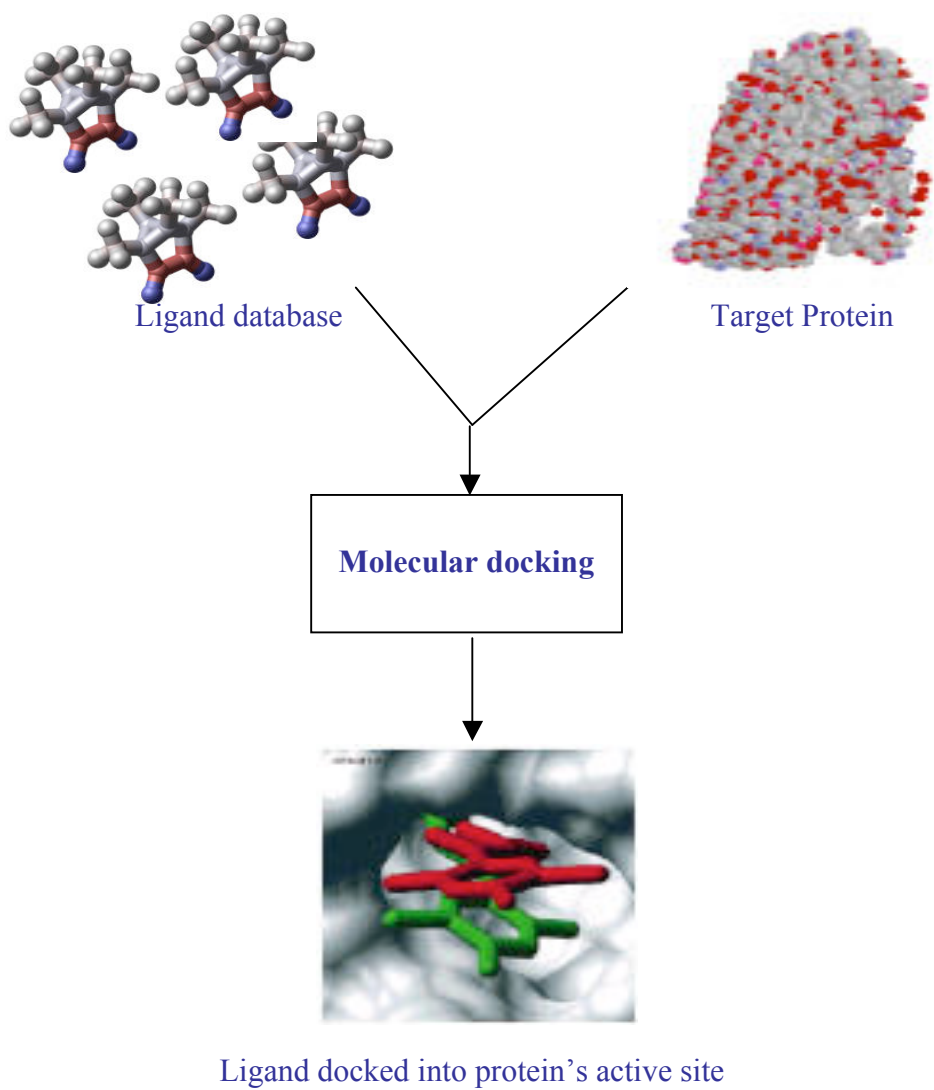

Ligand database    Target Protein

**Molecular docking**

Ligand docked into protein's active site

### 1.6.3 APPLICATIONS OF DOCKING

A binding interaction between a small molecule ligand and an enzyme protein may result in activation or inhibition of the enzyme. If the protein is a receptor, ligand binding may result in agonism or antagonism. Docking is most commonly used in the field of drug design— most drugs are small organic molecules, and docking may be applied to:

- Hit identification - docking combined with a scoring function can be used to quickly screen large databases of potential drugs insilico to identify molecules that are likely to bind to protein target of interest (see virtual screening).
- Lead optimization - docking can be used to predict in where and in which relative orientation a ligand binds to a protein (also referred to as the binding mode or pose). This information may in turn be used to design more potent and selective analogs.

### 1.7 REVERSE DOCKING

Ligand-protein docking has been developed and used in facilitating new drug discoveries. In this approach, docking single or multiple small molecules to a receptor site is attempted to find putative ligands. A number of studies have shown that docking algorithms are capable of finding ligands and binding conformations at a receptor site close to experimentally determined structures. These algorithms are expected to be equally applicable to the identification of multiple proteins to which a small molecule can bind or weakly bind. We introduce a ligand-protein inverse-docking approach for finding potential protein targets of a small molecule by the computer-automated docking search of a protein cavity database. This database is developed from protein structures in the Protein Data Bank (PDB). Docking is conducted with a procedure involving multiple-conformer shape-matching alignment of a molecule to a cavity followed by molecular-mechanics torsion optimization and energy minimization on both the molecule and the protein residues at the binding region. Scoring is conducted by the evaluation of molecular-mechanics energy and, when applicable, by the further analysis of binding competitiveness against other ligands that bind to the same receptor site in at least one PDB entry. The application of this approach may facilitate the prediction of unknown and secondary therapeutic target proteins and those related to the side effects and toxicity of a drug or drug candidate.

## 1.8 THE SEARCH ALGORITHM

The search space consists of all possible orientations and conformations of the protein paired with the ligand. With present computing resources, it is impossible to exhaustively explore the search space—this would involve enumerating all possible distortions of each molecule (molecules are dynamic and exist in an ensemble of conformational states) and all possible rotational and translational orientations of the ligand relative to the protein at a given level of granularity. Most docking programs in use account for a flexible ligand, and several are attempting to model a flexible protein receptor. Each "snapshot" of the pair is referred to as a pose. There are many strategies for sampling the search space. Here are some examples:

- Use a coarse-grained molecular dynamics simulation to propose energetically reasonable poses
- Use a "linear combination" of multiple structures determined for the same protein to emulate receptor flexibility
- Use a genetic algorithm to "evolve" new poses that are successively more and more likely to represent favorable binding interactions.

## 1.9 THE SCORING FUNCTION

The scoring function takes a pose as input and returns a number indicating the likelihood that the pose represents a favorable binding interaction.

Most scoring functions are physics-based molecular mechanics force fields that estimate the energy of the pose; a low (negative) energy indicates a stable system and thus a likely binding interaction. An alternative approach is to derive a statistical potential for interactions from a large database of protein-ligand complexes, such as the Protein Data Bank, and evaluate the fit of the pose according to this inferred potential.

There are a lot of structures from X-ray diffraction for complexes between proteins and high affinity ligands, but very few for low affinity ligands as these do not stay bound for long enough to be seen. Scoring functions trained with this data can dock high affinity ligands correctly, but they will also give plausible docked conformations for ligands that really are inactive. This gives a large number of false positive hits, i.e., ligands predicted to bind to the proteins that actually don't when placed together in a test tube.

One way to reduce the number of false positives is to recalculate the energy of the top-hit poses using a higher resolution (and therefore slow) technique like Generalized Born or Poisson-Boltzmann methods. However, typically the researcher will screen a database of tens to hundreds of thousands of compounds and test the top 60 or so in vitro, and to identify any true binders is still considered a success.

### 1.10  G-BIND:

It represents the free energy of binding, Gbind. The G-bind value has to be low for a structure to be stable. ΔΔG = -RTlnK. Binding (free) energy refers to that change in (free) energy for the following reaction:

Protein (in water) + ligand (in water) ----> protein-ligand complex (in water)

One factor that can strongly influence predicted free energy of binding is the ionization state of functional groups on the ligands and at the binding site at which calculations are performed.

### 1.11  RMS:

RMS refers to the Root Mean Squared Distance between the initial and final position of the ligand. The overall root mean square9RMS) deviation expression of any target molecule composed of m atoms in the training set may be written as a function of the parameters and geometry.

Rms=f(p1,p2,p3,……..,pm,x1,y1,z1,x2,y2,z2,….,xm,ym,zm)

P1,p2,..,pm represent the complete set of force field parameters required for the molecular mechanics calculations of the structure.

X1,y1,z1,….,xm,ym,zm are the optimized Cartesian coordinates

## 1.12 LOG P

The logarithm of the ratio of the concentrations of the un-ionized solute in the solvents is called log P. Hydrophobicity is represented by LogP. Partition coefficient is the ratio of concentrations of a compound in the two phases of a mixture of two immiscible solvents at equilibrium. Hence it is a measure of differential solubility of the compound between these two solvents. Normally one of the solvents chosen is water while the second is hydrophobic such as octanol. Partition coefficients are useful in estimating the distribution of drugs within the body. Hydrophobic drugs with high partition coefficients are preferentially distributed to hydrophobic compartments such as lipid bilayers of cells while hydrophilic drugs (low partition coefficients) preferentially are found in hydrophilic compartments such as blood serum.

## 1.13 LOGS

The aqueous solubility of a compound is denoted as logs. It significantly affects its absorption and distribution characteristics. Typically, a low solubility goes along with a bad absorption and therefore the general aim is to avoid poorly soluble compounds.

## 1.14 pKd:

The binding constant, pKd, is the negative logarithm of the inhibition constant Ki. The inhibitor constant, Ki, is an indication of how potent an inhibitor is; it is the concentration required to produce half maximum inhibition

Drug distribution within the body is determined mainly by free (unbound) concentration of drug in circulating plasma. The unbound fraction, in turn, depends on drug absorption by plasma proteins. Human Serum Albumin (HSA) is the most abundant blood plasma protein and is produced in the liver. HSA binding constants obtained by docking the molecule to both of the HSA active sites (Sudlow site I and Sudlow site II) are termed as Site1 pKd and site2 pKd respectively.

## 1.15  ALBUMIN PKD

Drug distribution within the body is determined mainly by free (unbound) concentration of drug in circulating plasma. The unbound fraction, in turn, depends on drug absorption by plasma proteins. Human Serum Albumin (HSA) is the most abundant blood plasma protein and is produced in the liver. HSA normally constitutes about 60% of human plasma protein. HSA concentrations in blood plasma range from 3.5 to 5.0g/l. It has been shown to shuttle a broad range of endogenous and exogenous ligands, including more than 70% of drugs.

Binding of a drug to HSA results in an increased solubility in plasma, decreased toxicity, and /or protection against oxidation of the bound ligand. Binding can also have a significant impact on the pharmacokinetics of drugs. Q-Albumin software takes a molecular structure and calculates HSA binding constant by docking the molecule to both of the HSA active sites (Sudlow site I and Sudlow site II).

## 1.16  DRUGLIKENESS

Druglikeness may be defined as a complex balance of various molecular properties and structure features which determine whether particular molecule is similar to the known drugs. These properties, mainly hydrophobicity, electronic distribution, hydrogen bonding characteristics, molecule size and flexibility and presence of various pharmacophoric features influence the behavior of molecule in a living organism, including bioavailability, transport properties, affinity to proteins, reactivity, toxicity, metabolic stability and many others.

## 1.17 LIPINSKI'S RULE

Lipinski's Rule of Five is a rule of thumb to evaluate druglikeness. The rule states, that most "drug-like" molecules have logP <= 5, molecular weight <= 500, number of hydrogen bond acceptors <= 10, and number of hydrogen bond donors <= 5. Molecules violating more than one of these rules may have problems with bioavailability. The rule is called "Rule of 5", because the border values are 5, 500, 2*5, and 5. The rule was formulated by Christopher A. Lipinski in 1997, based on the observation that most medication drugs are relatively small and lipophilic molecules. The rule describes molecular properties important for a drug's pharmacokinetics in the human body, including their absorption, distribution, metabolism, and excretion ("ADME"). However, the rule does not predict if a compound is pharmacologically active. The modification of the molecular structure often leads to drugs with higher molecular weight, more rings, more rotatable bonds, and a higher lipophilicity

## 1.18 HDAC-8 AS AN ANTI CANCER TARGET

Histone deacetylase (HDAC) and histone acetyltransferase (HAT) are enzymes that influence transcription by selectively deacetylating or acetylating the ε-amino groups of lysine located near the amino termini of core histone proteins. Chromatin acetylation correlates with transcriptional activity (euchromatin), whereas deacetylation correlates with gene silencing. HDACs are also involved in the reversible acetylation of non-histone proteins. Altered HDAC and/or HAT activities are present in many types of cancers.

Mammalian HDACs have been classified into three classes. Class I (HDACs 1, 2, 3 & 8; each of which contains a deacetylase domain exhibiting from 45% to 93% identity in amino acid sequence) are homologs of yeast RPD3 and localize to the nucleus. Class II (HDACs 4, 5, 6, 7, 9 & 10) are homologs of yeast Hda1 and are found in both the nucleus and cytoplasm. The molecular weights of which are all about two fold larger than those of the class I members, and the deacetylase domains are present within the C-terminal regions, except that HDAC-6 contains two copies of the domain, one within each of the N-terminal and C-terminal regions. HDAC11 has properties of both class I and class II HDACs. Class III (Sirt1 - Sirt7) are homologs of yeast Sir2 and form a structurally distinct class of NAD-dependent enzymes found in both the nucleus and cytoplasm.

Conserved from yeast to human, HDAC classes I and II are inhibited by trichostatin A (Prod. No. 380-068) and appear to use a divalent zinc-binding motif. The metal-coordinated active site activates an H2O molecule for direct targeting and hydrolysis of the acetyl group to form acetate. Acetylation of lysines in histones neutralizes the positive electric charge between the negatively charged DNA backbone and tips the balance towards relaxing the chromatin, while deacetylation would shift the balance back to condensing the chromatin and silencing gene expression. In a similar way PARP-1 adds to histones hundreds of negatively charged ADP-ribose units, which repel histones away from the negatively charged DNA backbone and thus induces chromatin relaxation to facilitate accession of DNA repair enzymes and gene expression.

HDAC inhibitors represent a relatively new group of targeted anticancer compounds, which are showing significant promise as agents with activity against a broad spectrum of neoplasms, at doses that are well tolerated by cancer patients. A number of small molecule inhibitors of HDAC, such as naturally occurring Trichostatin A (TSA), as well as synthetic compounds such as Suberoylanilide hydroxamic acid (SAHA), Scriptaid, Oxamflatin etc have been reported to induce differentiation of several cancer cell lines and suppress cell proliferation. But most of the inhibitors developed till date, including TSA and SAHA are derivatives of hydroxamic acid and are associated with poor pharmacokinetics and severe toxicity. They do not discriminate well among HDAC isozymes. Thus, there is a considerable interest in developing new non-hydroxamate HDAC inhibitors with few side effects.

Inhibitors of HDAC classes I and II emerge as potent anti-cancer agents. A proposed mechanism for the anti-tumor effects of HDAC inhibitors is that the accumulation of acetylated histones leads to activation (and repression) of the transcription of a selected number of genes whose expression causes inhibition of tumor cell growth and induction of apoptosis.

Disruption of HDACs has been linked to a wide variety of human cancers. HDAC inhibitors have been shown to be potent inducers of growth arrest, differentiation, and/or apoptotic cell death. Some newly synthesized compounds are potentially effective agents for cancer therapy and, possibly, cancer chemoprevention.

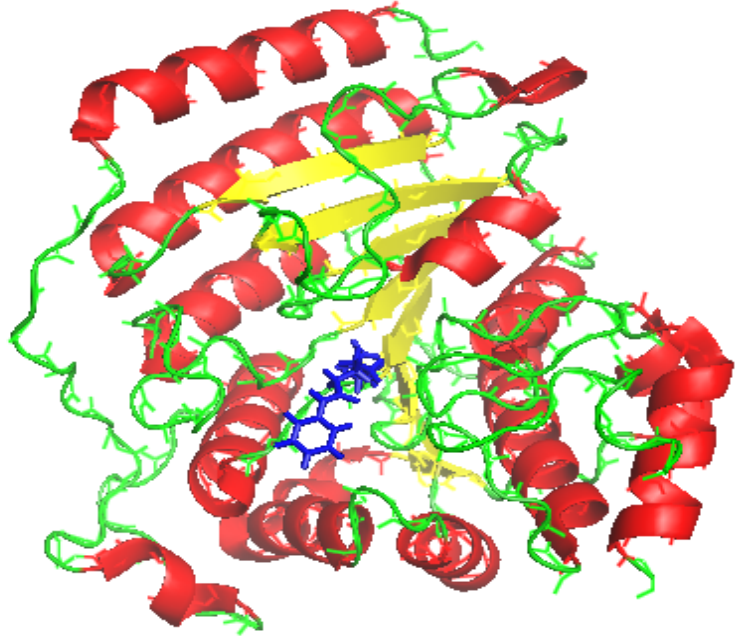

Figure-1: Structure of HDAC-8 with SAHA (in blue).

### 1.18.1 Classes of HDACs in higher eukaryotes

HDACs are classified in four classes depending on sequence identity and domain organization.

- Class I
    
    HDAC1, HDAC2, HDAC3, HDAC8
- Class II
    
    HDAC4, HDAC5, HDAC6, HDAC7A, HDAC9, HDAC10
- Class III
    
    Homologs of Sir2 in the yeast Saccharomyces cerevisiae
    
    Sirtuins in mammals (SIRT1, SIRT2, SIRT3, SIRT4, SIRT5, SIRT6, SIRT7)
- Class IV
    
    HDAC11

### 1.18.2 TRICHOSTATIN A

Trichostatin A is an organic compound that serves as an antifungal antibiotic and selectively inhibits the mammalian histone deacetylase family of enzymes. TSA inhibits the eukaryotic cell cycle during the beginning of the growth stage. TSA can be used to alter gene expression by interfering with the removal of acetyl groups from histones and therefore altering the ability of DNA transcription factors to access the DNA molecules inside chromatin. Thus, TSA has some uses as an anti-cancer drug. By promoting the expression of apoptosis-related genes, it may lead to cancerous cells surviving at lower rates, thus slowing the progression of cancer. Trichostatin A is harmful by inhalation, in contact with skin and if swallowed irritating to eyes, respiratory system and skin. Trichostatin A may cause sensitization by skin contact.

### 1.18.3 SUBEROYLANILIDE HYDROXAMIC ACID (SAHA)

Suberoylanilide hydroxamic acid (SAHA) is a novel histone deacetylase inhibitor with high potency in inducing differentiation of cultured murine erythroleukemia cells. SAHA induces cell cycle arrest and apoptosis in human breast cancer cells. SAHA reduces glioma progression in the organotypic brain environment.

### 1.18.4 SCRIPTAID

A novel histone deacetylase inhibitor, Scriptaid, enhances expression of functional estrogen receptor α (ER) in ER negative human breast cancer cells in combination with 5-aza 2′-deoxycytidine. The use of scriptaid resulted in a >100-fold increase in histone acetylation (Fig. 4) in cultured cells, which confirmed scriptaid as a HDAC inhibitor.

### 1.18.5 OXAMFLATIN

Oxamflatin is a novel antitumor compound that inhibits mammalian histone deacetylase. Oxamflatin caused an elongated cell shape with filamentous protrusions as well as arrest of the cell cycle at the G1 phase in HeLa cells. These phenotypic changes of HeLa cells were apparently similar to those by trichostatin A (TSA), a specific inhibitor of histone deacetylase (HDAC). Oxamflatin, like TSA, inhibited intracellular HDAC activity, as a result of which marked amounts of acetylated histone species accumulated. Oxamflatin induced the morphological changes of human cell lines characteristic of cells treated with TSA and other HDAC inhibitors during the experiments of in vitro cytotoxicity. Most of the inhibitors developed till date, including TSA and SAHA are derivatives of hydroxamic acid and are associated with poor pharmacokinetics and severe toxicity. They do not discriminate well among HDAC isozymes.

# AIM AND SCOPE

To identify novel potential ligands for HDAC-8 from a set of virtually screened molecules.

To calculate the drug likeness and IC50 of the novel potential ligands.

To optimize the novel potential ligands.

HDAC inhibitors have been shown to be potent inducers of growth arrest, differentiation, and/or apoptotic cell death. HDAC inhibitors also represent a relatively new group of targeted anticancer compounds, which are showing significant promise as agents with activity against a broad spectrum of neoplasms, at doses that are well tolerated by cancer patients. A number of small molecule inhibitors of HDAC, such as naturally occurring Trichostatin A (TSA), as well as synthetic compounds such as Suberoylanilide hydroxamic acid (SAHA), Scriptaid, Oxamflatin etc have been reported to induce differentiation of several cancer cell lines and suppress cell proliferation. But most of the inhibitors developed till date, including TSA and SAHA are derivatives of hydroxamic acid and are associated with poor pharmacokinetics and severe toxicity. They do not discriminate well among HDAC isozymes. Thus, there is a considerable interest in developing new non-hydroxamate HDAC inhibitors with few side effects.

# MATERIALS AND METHODS

**MATERIALS**

**3.1    PROTEIN DATA BANK** (http://www.rcsb.org)

The Protein Data Bank (PDB) is a repository for 3-D structural data of proteins and nucleic acids. These data typically obtained by X-ray crystallography or NMR spectroscopy and submitted by biologists and biochemists from around the world, are released into the public domain, and can be accessed for free.

The database contained 39,051 released atomic coordinate entries (or "structures"), 35,767 of that proteins, the rest being nucleic acids, nucleic acid-protein complexes, and a few other molecules. About 5,000 new structures are released each year. Data are stored in the mmCIF format specifically developed for the purpose.

The database stores information about the exact location of all atoms in a large biomolecule (although, usually without the hydrogen atoms, as their positions are more of a statistical estimate); if one is only interested in sequence data, i.e. the list of amino acids making up a particular protein or the list of nucleotides making up a particular nucleic acid, the much larger databases from Swiss-Prot and the International Nucleotide Sequence Database Collaboration should be used.

The structural data can be used to visualize the biomolecules with appropriate software, such as VMD, RasMol, PyMOL, Jmol, MDL Chime, QuteMol, web browser VRML plugin or any web-based software designed to visualize and analyze the protein structures such as STING. A recent desktop software addition is Sirius. The RCSB PDB website also contains resources for education, structural genomics, and related software.

## 3.2    SWISS PDB VIEWER (http://www.expasy.org/spdbv/)

DeepView - Swiss-PdbViewer is an application that provides a user friendly interface allowing to analyze several proteins at the same time. The proteins can be superimposed in order to deduce structural alignments and compare their active sites or any other relevant parts. Amino acid mutations, H-bonds, angles and distances between atoms are easy to obtain. DeepView - Swiss-PdbViewer has been developed by Nicolas Guex (GlaxoSmithKline R&D). Swiss-PdbViewer is tightly linked to SWISS-MODEL, an automated homology modeling server developed within the Swiss Institute of Bioinformatics (SIB) at the Structural Bioinformatics Group at the Biozentrum in Basel.

Working with these two programs greatly reduces the amount of work necessary to generate models, as it is possible to thread a protein primary sequence onto a 3D template and get an immediate feedback of how well the threaded protein will be accepted by the reference structure before submitting a request to build missing loops and refine side chain packing.

Swiss-PdbViewer can also read electron density maps, and provides various tools to build into the density. In addition, various modeling tools are integrated and command files for popular energy minimization packages can be generated.

## 3.3    PYMOL (www.pymol.org)

PyMOL is a user-sponsored molecular visualization system on an open-source foundation. It was created by Warren Lyford Delano and commercialized by Delano Scientific LLC, which is a private software company dedicated to creating useful tools that become universally accessible to scientific and educational communities. It is well suited to producing high quality 3D images of small molecules and biological macromolecules such as proteins. PYMOL is one of few open source visualization tools available for use in structural biology. The Py portion of the software's name refers to the fact that it extends, and is extensible by, the Python programming language.

### 3.4    ARGUSLAB (http://www.planaria-software.com)

Arguslab is a free program for calculating the docking modes of small molecules into protein binding sites. Arguslab is a molecular modeling program for windows 95/98 system. It consists of a user interface that supports OpenGL graphics display of molecular structure and runs quantum mechanical calculations using the Argus compute server. Arguslab contains an interactive 3D molecules builder that allows the user to build and manipulate complex structures and a rich suite of computational methods, both quantum mechanical and molecular mechanical for calculating ground or excited states properties.

### 3.5    PDB SUM (www.ebi.ac.uk/pdbsum)

PDB sum provides summaries and analyses all the structures in the PDB. Each summary gives an at-a-glance overview of the contents of a PDB entry in the terms of resolution and R-factor, number of protein chains, ligands, metal ions, secondary structure, fold cartoons and ligand interactions, etc. This is vital, not only for visualizing the structures concealed in PDB file, but also for drawing together in a single resource information at the 1D (sequence), 2D (motif) and 3D (structure) levels.

### 3.6    CHEMSKETCH

Advanced Chemistry Development, Inc., (ACD/Labs) has developed such an interface, and has integrated it with every desktop software module they produce. To date, over 800,000 chemists have incorporated ACD/Labs' chemical drawing and graphics package, ACD/ChemSketch, into their daily routines. Academic institutions worldwide have adopted this software as an interactive teaching tool to simplify and convey chemistry concepts to their students, and publishing bodies such as Thieme, the publisher of Science of Synthesis, consider it to be "...supportive of the organic chemistry publisher's role, both in the construction of compounds and their basic analysis."

ACD/ChemSketch is an advanced chemical drawing tool and is the accepted interface for the industry's best NMR and molecular property predictions, nomenclature, and analytical data handling software.

ACD/ChemSketch is also available as freeware, with functionalities that are highly competitive with other popular commercial software packages. The freeware contains tools for 2D structure cleaning, 3D optimization and viewing, InChI generation and conversion, drawing of polymers, organometallics, and Markush structures—capabilities that are not even included in some of the commercial packages from other software producers. Also included is an IUPAC systematic naming capability for molecules with fewer than 50 atoms and 3 rings. The capabilities of ACD/ChemSketch can be further extended and customized by programming.

The commercial version of ACD/ChemSketch offers additional capabilities above and beyond the freeware offering. It includes a number of advanced features including a dictionary of more than 158,000 trivial, common, and trade names with their corresponding structures. It allows the user to view SDfiles, and search Microsoft Word or Adobe PDF reports, SDfiles, molfiles, and CambridgeSoft ChemDraw files by chemical structure, substructure, or structure similarity.

## 3.7   CORINA

CORINA is a program for the fast and efficient generation of high-quality three-dimensional molecular models.
Corina is a rule and data based program system, that automatically generated three dimensional atomic coordinates from the constitution of a molecule as expressed by a connection table or linear code, and which is powerful and reliable to convert large databases of several hundreds of compounds.  Corina is applicable to the range of organic chemistry.  Structures, which can be expressed in a valence bond notation can be processed.  It does not provide any upper limit to the size of the ring system. The program fully considers stereo chemical information and generates the defined stereo isomer.  Corina processes structures containing atoms with up to six neighbors; thus, even metal complexes can be processed.  It generates one low energy conformation for each input structure. For ring system consisting of up to nine atoms, multiple conformation can be generated- a useful feature for building flexible 3- D databases. The program automatically detects stereo centers (tetrahedral centers and Cis/Trans double bonds) and is able to generate all possible isomers. Duplicate isomers, such as meso compounds are identified and removed as well as geometrically strained configurations. Corina can process a variety of standard file format for structure input and output (e.g.: MDL SD/ RD FILE, SMILES, SYBYL MOLFILE and MOL2, PDB, Macro Model, Maestro or CIF).

Corina delivers structures of high quality. The RMS deviation of corina built models from published X-Ray structures is among the best of all commercially available conversion programs. It is fast, robust and provides excellent conversion rate. Corina offers many features to influence the 3- D generation process. It provides an interface to the ligand docking program FlexX.

### 3.8  PUBCHEM

PubChem is a database of chemical molecules. The system is maintained by the National Center for Biotechnology Information (NCBI), a component of the National Library of Medicine, which is part of the United States National Institutes of Health (NIH). PubChem can be accessed for free through a web user interface. PubChem contains substance descriptions and small molecules with fewer than 1000 atoms and 1000 bonds. PubChem contains its own online molecule editor with SMILES/SMARTS and InChI support that allows the import and export of all common chemical file formats to search for structures and fragments. Each hit provides information about synonyms, chemical properties, chemical structure including SMILES and InChI strings, bioactivity, and links to structurally related compounds and other NCBI databases like PubMed. PubChem also provides a fast chemical structure similarity search tool.

PubChem Compound: Search unique chemical structures using names, synonyms or keywords. Links to available biological property information are provided for each compound.

PubChem Substance: Search deposited chemical substance records using names, synonyms or keywords. Links to biological property information and depositor web sites are provided.

PubChem BioAssay: Search bioassay records using terms from the bioassay description, for example "cancer cell line". Links to active compounds and bioassay results are provided.

Structure Search: Search PubChem's Compound database using a chemical structure as the query. Structures may be sketched or specified by SMILES, MOL files, or other formats.

### 3.9 C@ROL

C@ROL (Compound Access & Retrieval On Line) is a web-based warehouse for chemical compounds. C@rol stores the 2D and 3D structures as well as multiple conformation of a molecule. It reads different file formats for chemical structure registration. C@rol exports structures and data in various file formats.

It has two major applications.

Chemical and pharmaceutical companies can merge the products offered by various suppliers into one system in order to simplify and speed up the retrieval and in-house ordering process of chemical compounds.

Suppliers of chemical compounds can present their products on the internet with full retrieval capabilities in a web-based application.

C@ROL provides various search capabilities including innovative similarity searches in its chemical databases. Thus, it makes it as easy as possible to find the best offer for your requirements.

#### *3.9.1 Features*

* Web-based graphical user interface.

*Graphical input of compounds by molecule editor.

*Various rapid search methods for chemical structures (Structure, substructure, similarity, transformation, and 3D pharmacophore searches).

*Search methods for chemical names and properties (IUPAC name, CAS registry number, molecular formula, molecular weight, etc.)

*Focusing the hits on a given synthesis problem or on a specific biological activity.

*Various output formats for chemical structures.

*Optional 3D visualization.

*Integrated email functionality

*Loading of databases of your choice

*Administration capabilities, e.g. user management

### 3.10   WL VIEWERLITE

The WebLab Viewer is an innovative software tool for examining the 3D structure of molecular models, and for communicating the resulting information with colleagues. With the WebLab Viewer, a molecule can be viewed as a wireframe, a stick model, a ball and stick model, or a space-filling model.  The model can be rotated, translated, or scaled to any particular viewpoint.  Distances, angles, torsions, and stereochemistry can be easily measured; these variables are instantly updated whenever the local geometry is modified.  You can color or label atoms to emphasize different attributes.

he WebLab Viewer reads all of the most popular molecular file formats.  In addition, you can paste molecules from popular 2D drawing packages such as ISIS Draw or ChemDraw into the Viewer.  Molecules drawn in 2D in ISIS Draw or ChemDraw are converted to the proper 3D geometry automatically when brought into the WebLab Viewer.  We can add or remove hydrogens and determine the R/S stereochemistry for chiral atoms.

The WebLab Viewer provides numerous options for advanced protein and DNA visualization. Hydrogen bonds can also be displayed. The WebLab Viewer can create PDB and MOL files for exporting molecular information to other applications, VRML files for display in VRML-compliant browsers, and JPEG, GIF, and BMP files for use as graphics.

## 3.11  VEGA ZZ

VEGA ZZ is the evolution of the well known VEGA OpenGL package and includes several new features and enhancements making your research jobs very easy. VEGA was originally developed to create a bridge between most of the molecular software packages only, but in the years, enhancing its features, it's evolved to a complete molecular modeling suite.

### *3.11.1  3D Features*:

Extreme OpenGL implementation for an incredible real-time rendering quality: lighting (4 customizable light sources + ambient light), alpha blending, hardware anti-aliasing, material management, 3D backgrounds.

Stereo view (shutter or anaglyphic glasses).

Hardware and software offline rendering.

3D molecule view: wireframe with multivector bonds, CPK, ball & stick, stick, trace and tube. All representations can be mixed thanks to the selection tool.

Atom labels.

Enhanced atom coloring methods.

Atom selection & picking.

3D surface: dotted, mesh, solid, solid transparent. Thanks to the Hyper Drive technology, the calculation is very fast. The surfaces can be colored by atom, residue, chain, segment, molecule, surface ID and property. The color gradient used in the property coloring mode can be customized by the user defining the number and the type of colors.

Multiple surface management.

All 3D objects can be managed with mouse, joystick and dials.

3D interactive monitors calculated in real time (distance, angle, torsion and angle between two planes).

Simulation trajectory visualization and animation.

Snapshot, hardware and software image rendering with the capability to create images bigger than the monitor size. The software rendering includes an anti-aliasing algorithm with user-selectable 4x or 16x super sampling. The supported output formats are: BMP, GIF 256 colors, JPEG, PCX, PNG, PNM, RAW, SGI, TGA and TIFF.

Vector graphic rendering engine. It's possible to export the view in PostScript, Encapsulated PostScript, PDF, LaTex, POV-Ray and VRML 2.0 formats.

## 3.12  GENETIC OPTIMIZATION LIGAND DOCKING (GOLD)

GOLD (Genetic Optimization for Ligand Docking) is a genetic algorithm for docking flexible ligands into protein binding sites. GOLD is an automated ligand docking program that uses a genetic algorithm to explore the full range of ligand conformational flexibility with partial flexibility of the protein, and satisfies the fundamental requirement that the ligand must displace loosely bound water on binding. Numerous enhancements and modifications have been applied to the original technique resulting in a substantial increase in the reliability and the applicability of the algorithm. The advanced algorithm has been tested on a dataset of 100 complexes extracted from the Brookhaven Protein Data Bank. When used to dock the ligand back into the binding site, GOLD achieved a 71% success rate in identifying the experimental binding mode.

GOLD provides all the functionality required for docking ligands into protein binding sites from prepared input files. GOLD will likely be used in conjunction with a modeling program since you will be required to create and edit starting models, e.g. add all hydrogen atoms, including those necessary for defining the correct ionization and tautomeric states of the residues. Commonly used molecular modeling environments include:

    SYBYL (http://www.tripos.com/)

    Insight II or Cerius2 (http://www.accelrys.com/).

Predicting how a small molecule will bind to a protein is difficult, and no program can guarantee success. The next best thing is to measure as accurately as possible the reliability of the program, i.e. the chance that it will make a successful prediction in a given instance. For that reason, GOLD has been tested on a large number of complexes extracted from the Protein Data Bank. The overall conclusion of these tests was that the top-ranked GOLD solution was correct in 70-80% of cases.

GOLD offers a choice of scoring functions, GoldScore, ChemScore and User Defined Score which allows users to modify an existing function or implement their own scoring function. With respect to using the GoldScore or ChemScore functions one may give a successful prediction where the other fails, but their overall success rates are about the same.

Different values of the genetic algorithm parameters may be used to control the balance between the speed of GOLD and the reliability of its predictions. GOLD will only produce reliable results if it is used properly and correct atom typing for both protein and ligand is particularly important.

GOLD may be used in serial or parallel modes.

GOLD will dock each ligand several times starting each time from a different random population of ligand orientations. The results of the different docking runs are ranked by fitness score.

The number of dockings to be performed on each ligand is set when the ligand file is defined.

By default the number of dockings to be performed on each ligand is 10.

The total time spent docking a ligand obviously depends on the number of docking runs, so you can make GOLD go faster by reducing this number. However, it is useful to perform at least a few docking runs on each ligand. This increases the chances of getting the right answer. Also, if the same answer is found in several different docking runs, it is usually a strong indicator that the answer is correct.

### *3.12.1 GOLD features:*

A genetic algorithm (GA) for protein-ligand docking.

Full ligand flexibility.

Partial protein flexibility, including protein side chain and backbone flexibility for up to ten user-defined residues.

Energy functions partly based on conformational and non-bonded contact information from the CSD.

A choice of GoldScore, ChemScore or Astex Statistical Potential (ASP) scoring functions.

Extensive options for customizing or implementing new scoring functions through a Scoring Function Application Programming Interface, allowing users to modify the GOLD scoring-function mechanism in order to: implement their own scoring function or enhance existing scoring functions.

Customize docking output.

A choice of GoldMine or SILVER for post-processing docking results.

Automatic consideration of cavity bound water molecules.

Improved handling and control of metal coordination geometries.

Options for generating diverse solutions, based on RMSD.

Automatic derivation of GA settings for particular ligands.

## 3.13 SILVER

SILVER is a program included for use with GOLD and can be used to post-process docking results for large numbers of ligands. SILVER allows easy set-up and calculation of a variety of customizable descriptors (parameters that describe dockings) that quantify, amongst other things.

The hydrogen-bonding interactions that occur between protein and docked ligand.

The H-bond interactions that do not occur, e.g. a protein H-bond donor that is prevented from forming a hydrogen bond by a ligand hydrophobic group.

Other close contacts between protein and ligand.

The buried surface area of the ligand, or of certain types of atoms in the ligand (e.g. hydrophobic atoms).

Whether particular regions of the binding site are occupied by the ligand.

Simple properties such as the number of H-bonding ligand atoms, molecular weight of ligand, number of rotatable bonds.

Although not its primary purpose, SILVER also serves as a browser for visualizing docking results from GOLD.

## 3.14 Q-ALBUMIN

Q-Albumin software takes a molecular structure and calculates HSA (Human Serum Albumin) binding constant by docking the molecule to both of the HSA active sites (Sudlow site I and Sudlow site II).

Drug distribution within the body is determined mainly by free (unbound) concentration of drug in circulating plasma. The unbound fraction, in turn, depends on drug absorption by plasma proteins. Human Serum Albumin (HSA) is the most abundant blood plasma protein and is produced in the liver.     Quantum Pharmaceuticals are the vendors of Q-Albumin software and is a commercially available one.

### 3.15  QUANTAMSOFTWARE

QUANTUM is a drug discovery and computational chemistry tool firmly based on fast ab initio molecular, quantum and statistical physics methods. The implemented algorithms allow to conduct calculations in complicated chemical environments including ions and hetgroups, with full protein and the drug candidates flexibility accounted. The core of QUANTUM is a docking tool useful for chemical library screening and identifying compounds with strong binding affinity to a given disease target.

QUANTUM employs quantum mechanics, thermodynamics, and an advanced continuous water model for solvation effects to calculate ligands binding affinities. This approach differs dramatically from scoring functions that are commonly used for binding affinity predictions. By including the entropy and aqueous electrostatics contributions in to the calculations directly, QUANTUM algorithms produce much more accurate and robust values of binding free energies.

Interaction of a ligand with a protein is characterized by the value of binding free energy. The free energy (F) is the thermodynamic quantity, that is directly related to experimentally measurable value of inhibition constant (IC50) and depends on electrostatic, quantum, aqueous solvation forces, as well as on statistical properties of interacting molecules.

QUAMTUM SOFTWARE is provided by Quantum pharmaceuticals and is a commercial one.

### 3.16 METHODOLOGY

The X-ray crystallographic structure of Human HDAC-8 complexed with SAHA was retrieved from Protein Data Bank (PDB ID: 1T69). This structure was saved as a standard PDB file. The ligand and the receptor protein are separated and saved in two different files using Swiss PDB viewer. The Smiles formula of ligand was retrieved from PDB and used to generate the original three dimensional structure of the ligand. For this the free standing molecular building tool CORINA is used. The active site of the protein was defined as the residues within 4Å vicinity of the ligand molecule with the help of PDB viewer. Preparation of active site involved correcting the ionization states of key amino acid side chains, adding hydrogen and listing out all the atoms making the active site as a text file. For this Argus Lab was used. GOLD (Genetic Optimization for Ligand Docking), was used for protein-ligand docking.

GOLD (Genetic Optimization for Ligand Docking), was used for protein-ligand docking. The CORINA generated ligand structure was docked into the active site and the GoldScore was recorded. Default settings were chosen and GOLD was run under Standard mode.

For screening of chemical databases, library screening settings was selected. Databases containing 3D structures of thousands of small molecules were downloaded in SDF format. The databases screened include CHEMBANK, KEGG, and NCTER which provided a total of around six thousand molecules.

Top ranking molecules from the best rank file were recorded. It was followed by strict visualization using Pymol and "hits" were selected. These molecules were then docked independently for GoldScore and then ChemScore. Molecules which got high scores for both GoldScore and ChemScore were selected. These selected structures were opened in chemsketch and functional groups were added to the ligands. These ligands were saved as pdb format. These ligands were then taken and again docked. Molecules which got high scores for both GoldScore and ChemScore were selected and compared with the previous results. Those with best results were then taken for calculating H-bond interactions and close contacts in silver. Bioactivity prediction of the best ranked ligands was done using Quantum software. Their physicochemical properties were also analyzed in Q-albumin software. Ultimately the results obtained were analyzed to predict the best potential ligand for HDAC-8.

# RESULTS AND DISCUSSION

**Table 4.1: Goldscore and chemscore of the known ligands**

| NAME | GOLDSCORE | CHEMSCORE |
|------|-----------|-----------|
| B3N  | 58.19     | 21.10     |
| SHH  | 66.06     | 20.90     |
| TSN  | 64.00     | 24.88     |

**Table 4.2: Goldscore and chemscore results of kegg ligands**

| Keg id | Name | Goldscore | Chemscore |
|--------|------|-----------|-----------|
| C00124 | Isopentenyl diphosphate; delta3-Isopentenyl diphosphate; delta3-Methyl-3-butenyl diphosphate | 60.79 | 6.77 |
| C00341 | Geranyl diphosphate | 63.33 | 11.82 |
| C00404 | (Phosphate)n; (Phosphate)n+1; (Phosphate)n-1 | 70.55 | 6.69 |
| C00536 | Triphosphate; Tripolyphosphate | 67.43 | 5.85 |
| C00677 | Deoxynucleoside triphosphate | 83.87 | 5.59 |
| C02569 | Neryl diphosphate; Neryl pyrophosphate | 64.59 | 11.56 |
| C03190 | (+)-Bornyl-diphosphate | 68.69 | 12.58 |
| C03279 | Inorganic triphosphate | 68.43 | 6.67 |
| C04093 | poly-cis-Polyprenyl diphosphate | 62.56 | 11.53 |
| C05308 | Linaloyl diphosphate | 61.72 | - |
| C05470 | Urocortisone | 62.20 | - |
| C05806 | Polyprenyl diphosphate | 64.07 | 11.19 |

**Table 4.3: Silver results of known ligands**

| Name | Close contact with protein residues | H-bonds |
|------|-------------------------------------|---------|
| B3N  | TYR306:OH | HIS180 TYR306 |
| SHH  | HIS180:ND1 HIS180:CE1 | HIS180 |

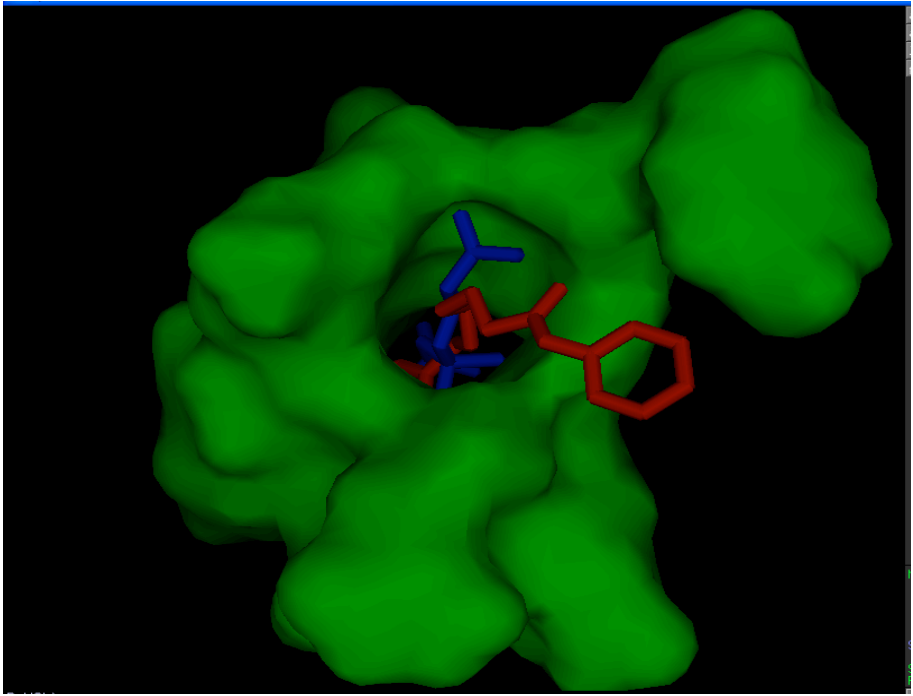

*Fig: 1: Isopentenyl diphosphate; delta3-Isopentenyl diphosphate; delta3-Methyl-3-butenyl diphosphate (kegg c00124)*

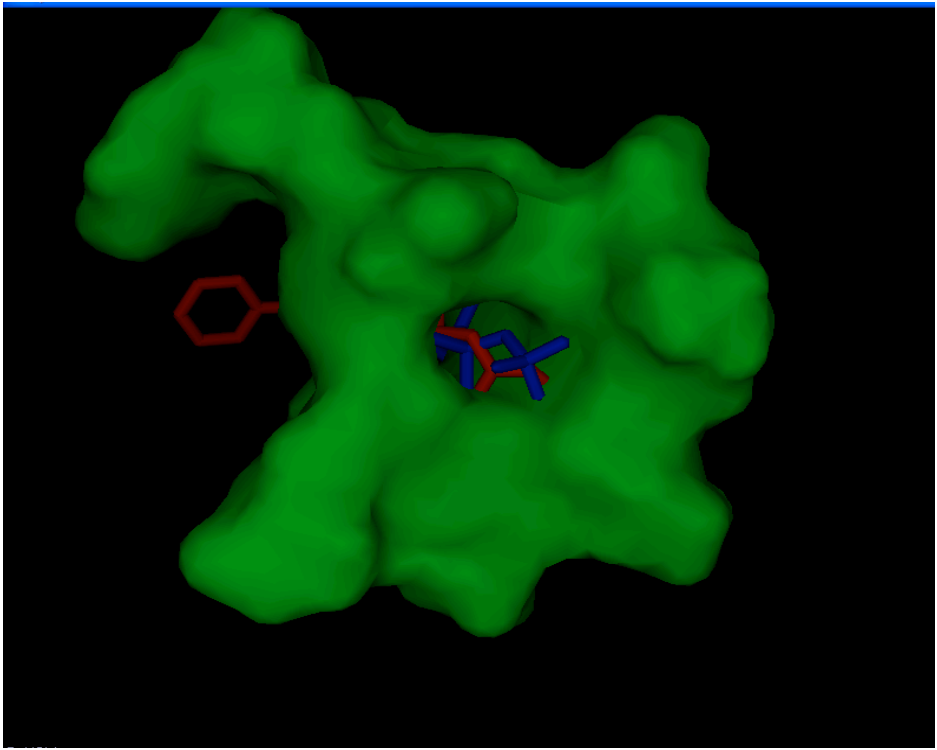

*Fig: 2: Geranyl diphosphate (kegg c00341)*

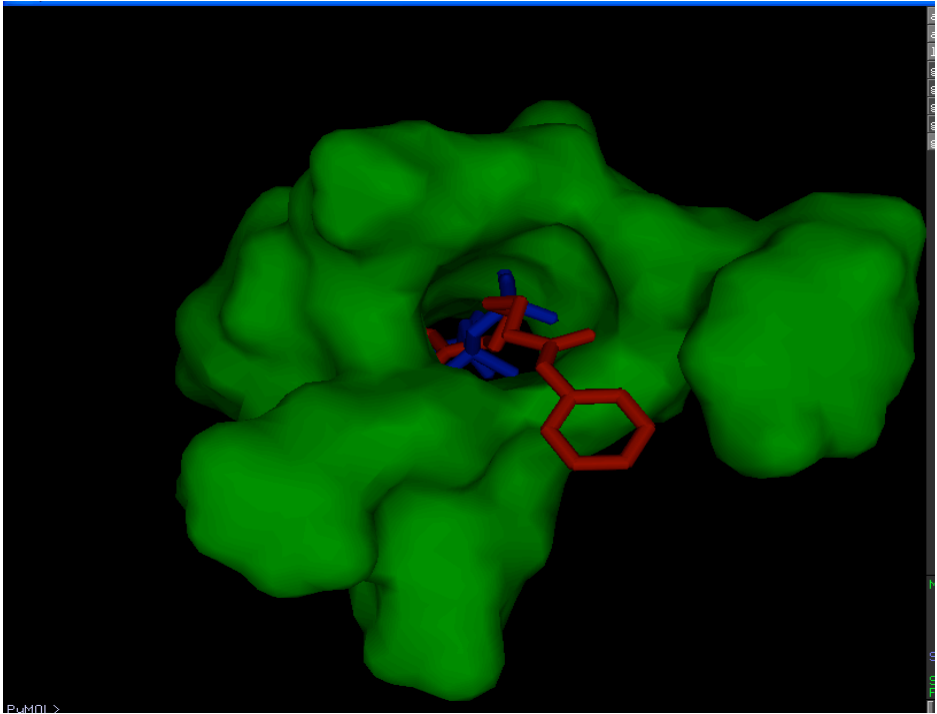

*Fig: 3: (Phosphate)n; (Phosphate)n+1; (Phosphate)n-1 (Kegg c00404)*

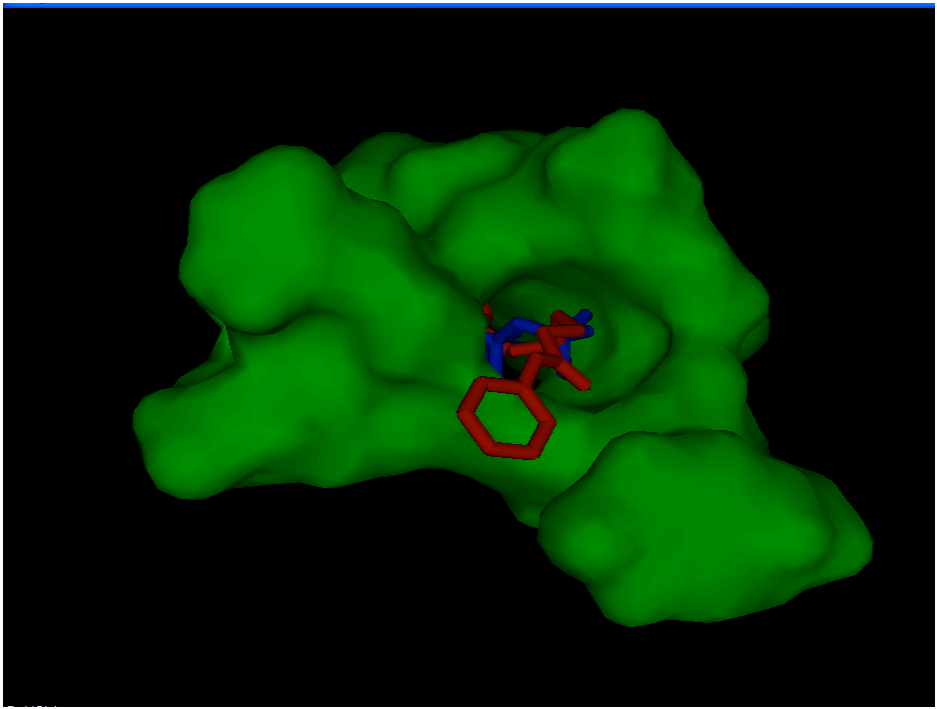

*Fig: 4: Triphosphate; Tripolyphosphate (kegg c00536)*

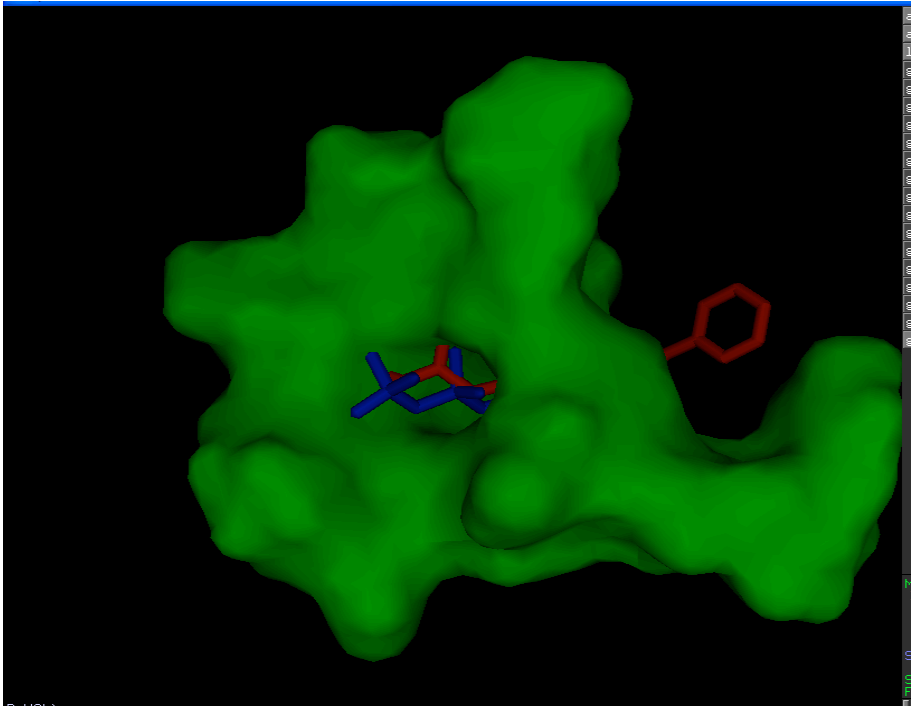

*Fig: 5: Deoxynucleoside triphosphate (kegg c00677)*

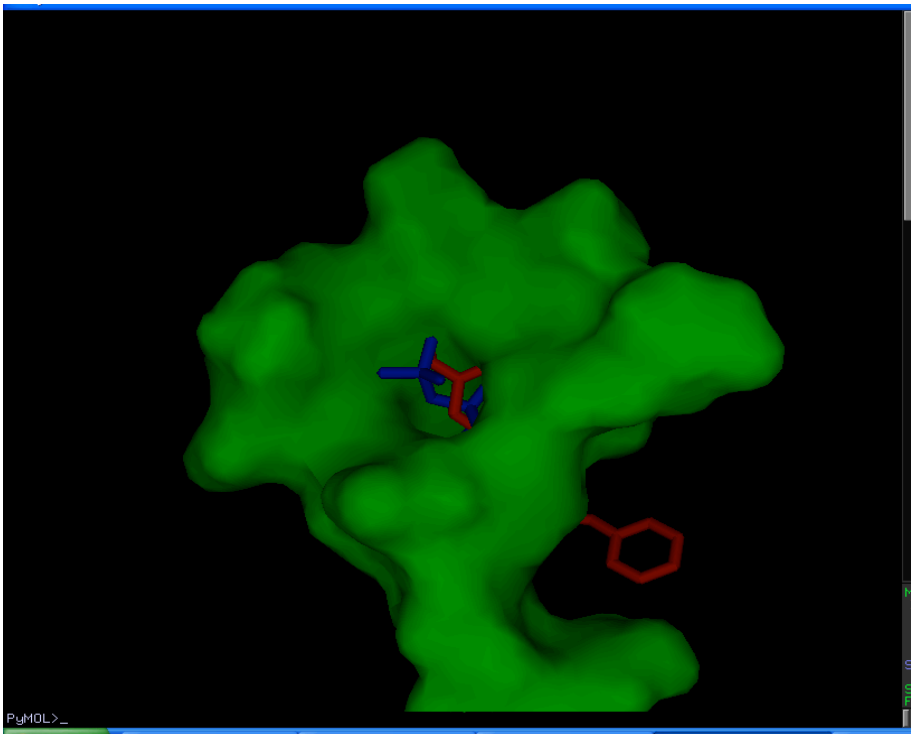

*Fig: 6: Neryl diphosphate; Neryl pyrophosphate (kegg c02569)*

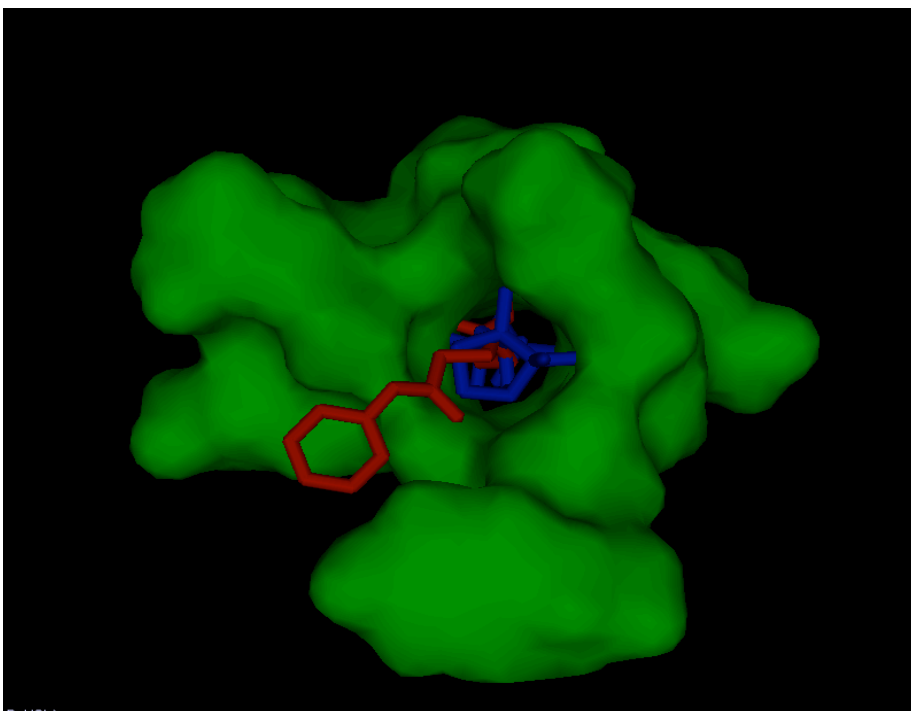

*Fig: 7: (+)-Bornyl-diphosphate (kegg c03190)*

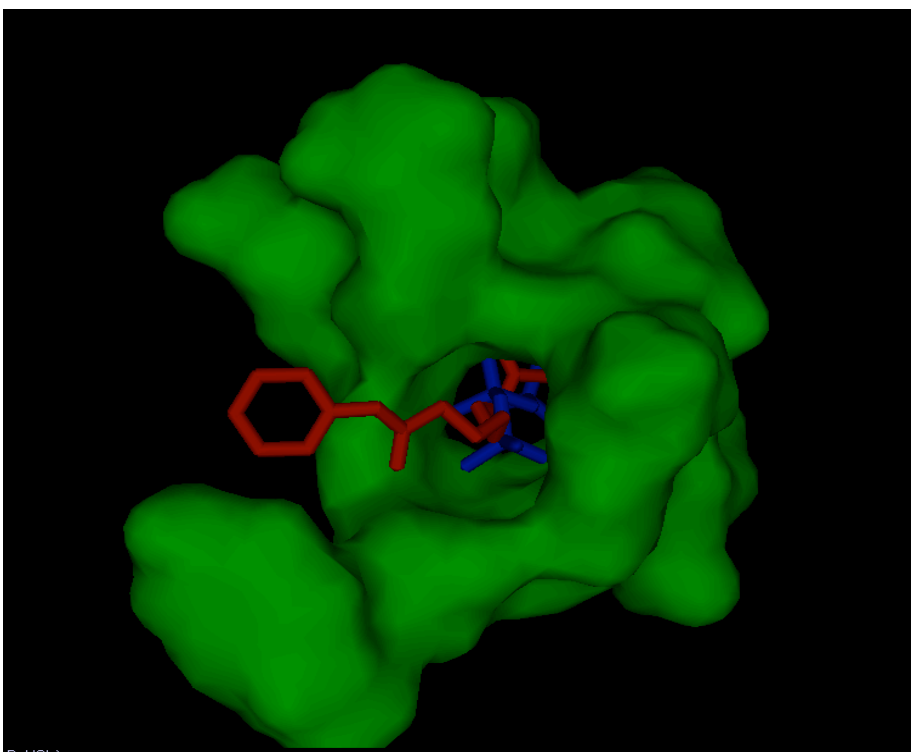

*Fig: 8: Inorganic triphosphate (kegg c03279)*

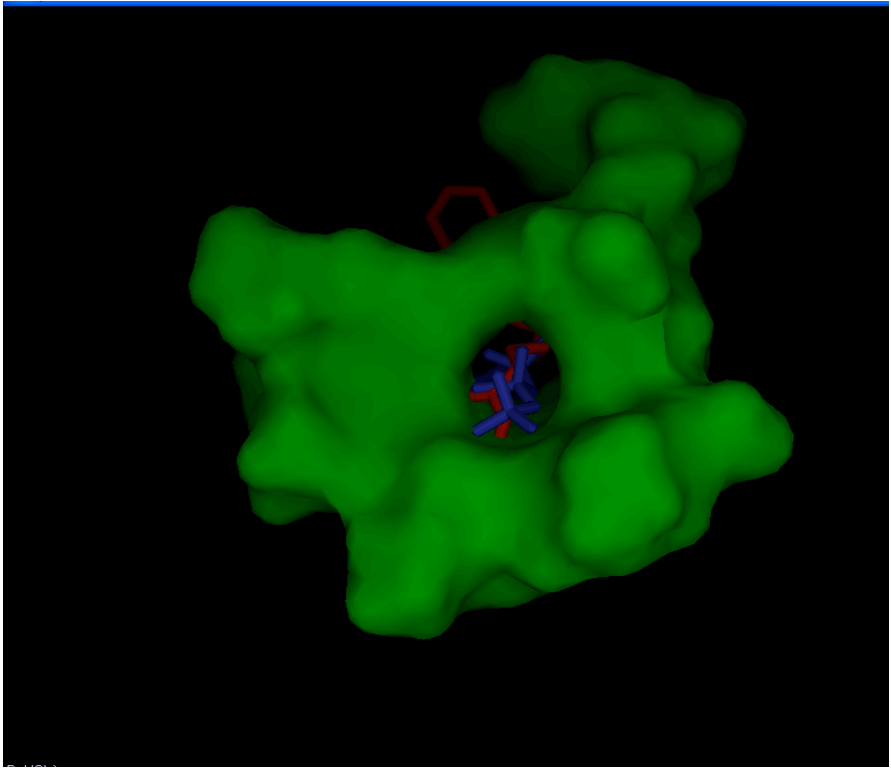

*Fig: 9: poly-cis-Polyprenyl diphosphate (kegg c04093)*

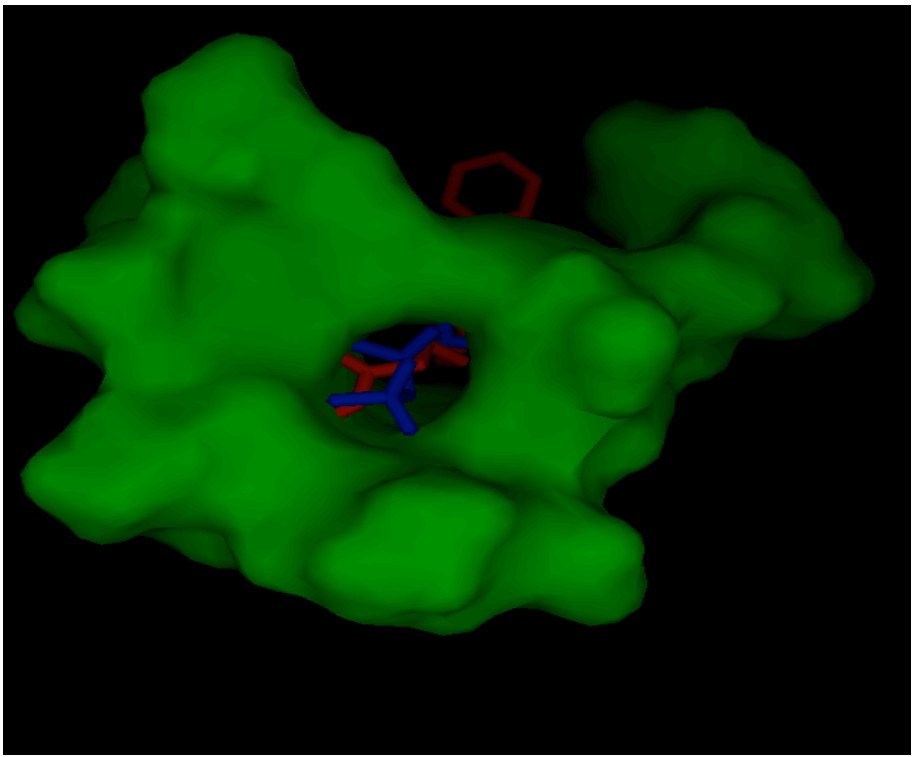

Fig: 10: Linaloyl diphosphate (kegg c05308)

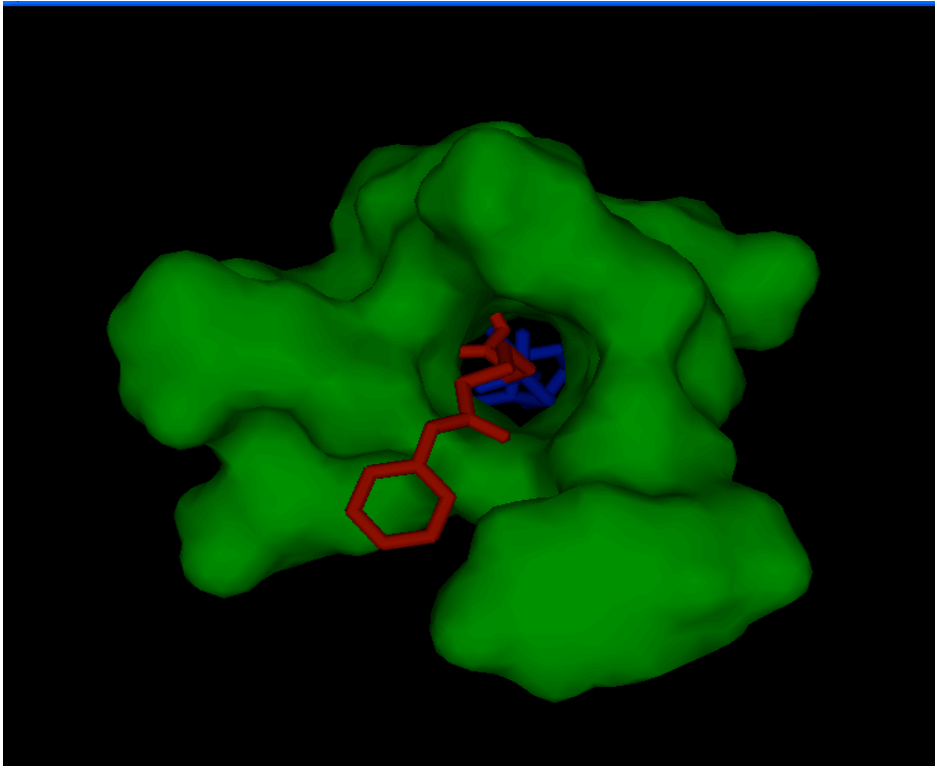

*Fig: 11: Urocortisone (kegg c05470)*

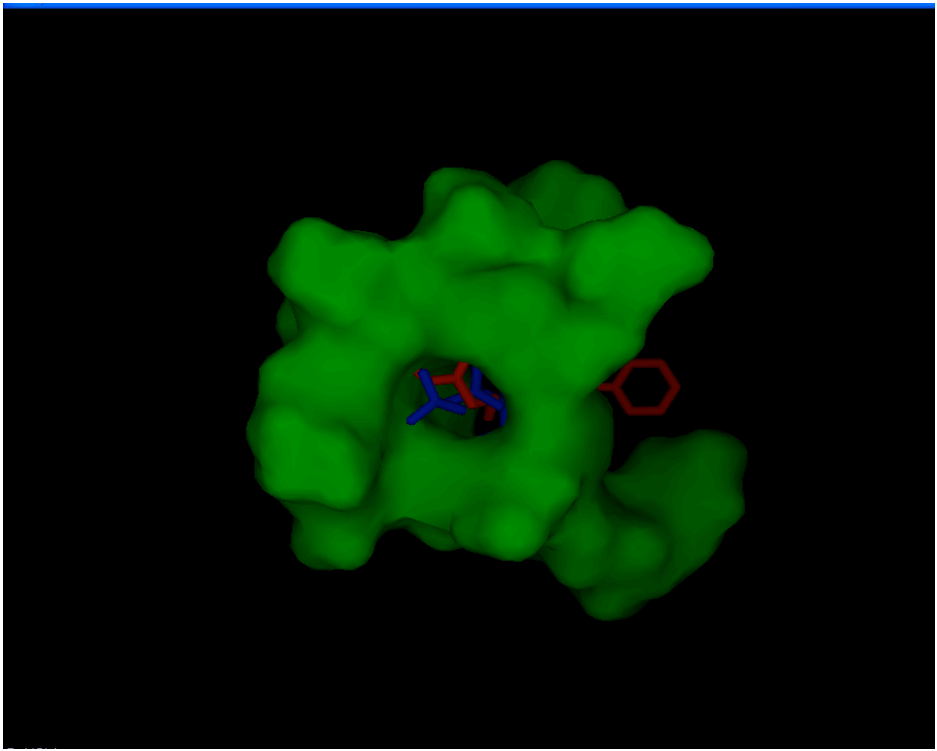

*Fig: 12: Polyprenyl diphosphate (kegg c05806)*

**Table 4.4: Silver results of ligands taken from kegg database**

| Name | Close contacts with protein residues | H-bonds |
|---|---|---|
| Isopentenyl diphosphate; delta3-Isopentenyl diphosphate; delta3-Methyl-3-butenyl diphosphate | HIS180:CE1<br>PHE152:CD2 | TYR306:H |
| Geranyl diphosphate | HIS142:CD2<br>HIS180:CE1<br>PHE152:CD2 | TYR306:H<br>HIS180:H |
| (Phosphate)n; (Phosphate)n+1; (Phosphate)n-1 | HIS143:CD2<br>PHE208:CE1<br>PHE152:CD2 | TYR306:H<br>HIS180:H |
| Triphosphate; Tripolyphosphate | PHE152:CD2<br>HIS143:CD2<br>PHE208:CE1<br>HIS180:CE1 | TYR306:H<br>HIS180:H |
| Deoxynucleoside triphosphate | HIS180:CE1<br>HIS142:NE2<br>HIS142:CD2<br>HIS143:CD2 | HIS142:H<br>TYR306:H<br>PHE208:H |
| Neryl diphosphate; Neryl pyrophosphate | HIS180:CE1<br>PHE152:CD2<br>PHE152:CD1<br>PHE152:CE1 | TYR306:H<br>HIS180:H |
| (+)-Bornyl-diphosphate | PHE152:CD2<br>HIS180:CE1<br>HIS143:CD2 | TYR306:H<br>HIS180:H |
| Inorganic triphosphate | PHE152:CD2<br>HIS143:CD2<br>HIS142:CD2<br>PHE208:CE1 | TYR306:H<br><br>HIS180:H |
| poly-cis-Polyprenyl diphosphate | PHE152:CD2<br>HIS180:CE1 | TYR306:H<br><br>HIS180:H |
| Linaloyl diphosphate | PHE152:CD2<br>HIS143:CD2<br>HIS142:CD2 | TYR306:H<br>HIS180:H |
| Urocortisone | HIS180:CE1<br>PHE152:CD2<br>PHE152:CG<br>GLY151:CA<br>GLY151:C | TYR306:H<br>HIS180:H<br>PHE208:H |
| Polyprenyl diphosphate | TYR306:CD2<br>TYR306:CE1<br>ASP178:CG | HIS180:H<br>ASP178:H<br>ASP267:H |

**Table 4.5: Druglikiness results of ligands taken from keg database.**

| Keg id | Name | LogP | Logs | Lipinski's rule |
|--------|------|------|------|-----------------|
| C00124 | Isopentenyl diphosphate; delta3-Isopentenyl diphosphate; delta3-Methyl-3-butenyl diphosphate | 0.8 | -1.0 | true |
| C00341 | Geranyl diphosphate | 1.5 | -2.0 | true |
| C00404 | (Phosphate)n; (Phosphate)n+1; (Phosphate)n-1 | -0.7 | 0.7 | true |
| C00536 | Triphosphate; Tripolyphosphate | -0.7 | 0.7 | true |
| C00677 | Deoxynucleoside triphosphate | - | - | - |
| C02569 | Neryl diphosphate; Neryl pyrophosphate | 2.5 | -3.2 | true |
| C03190 | (+)-Bornyl-diphosphate | 2.2 | -2.3 | true |
| C03279 | Inorganic triphosphate | - | - | - |
| C04093 | poly-cis-Polyprenyl diphosphate | 2.5 | -3.2 | true |
| C05308 | Linaloyl diphosphate | - | - | - |
| C05470 | Urocortisone | 1.8 | -3.5 | true |
| C05806 | Polyprenyl diphosphate | 2.5 | -3.2 | true |

**Table 4.6: Properties of ligands taken from kegg database**

| Keg id | Name | Mol.wt | H-bond donor count | H-bond acceptor count |
|---|---|---|---|---|
| C00124 | Isopentenyl diphosphate; delta3-Isopentenyl diphosphate; delta3-Methyl-3-butenyl diphosphate | 246.092102 g/mol | 3 | 7 |
| C00341 | Geranyl diphosphate | 314.209122 g/mol | 3 | 7 |
| C00404 | (Phosphate)n; (Phosphate)n+1; (Phosphate)n-1 | 257.954983 g/mol | 5 | 10 |
| C00536 | Triphosphate; Tripolyphosphate | 257.954983 g/mol | 5 | 10 |
| C00677 | Deoxynucleoside triphosphate | 291.6865 g/mol | 3 | 5 |
| C02569 | Neryl diphosphate; Neryl pyrophosphate | 314.209122 g/mol | 3 | 7 |
| C03190 | (+)-Bornyl-diphosphate | 314.209122 g/mol | 3 | 7 |
| C03279 | Inorganic triphosphate | 257.954983 g/mol | 5 | 10 |
| C04093 | poly-cis-Polyprenyl diphosphate | 314.209122 g/mol | 3 | 7 |
| C05308 | Linaloyl diphosphate | 314.209122 g/mol | 3 | 7 |
| C05470 | Urocortisone | 364.47578 g/mol | 3 | 5 |
| C05806 | Polyprenyl diphosphate | 314.209122 g/mol | 3 | 7 |

**Table 4.7: lc50 results of ligands taken from kegg database**

| Keg id | Name | lc50 | Gbind,KJ/ | RMS |
|---|---|---|---|---|
| C00124 | Isopentenyl diphosphate; delta3-Isopentenyl diphosphate; delta3-Methyl-3-butenyl diphosphate | 9.13e-002 | -6.05 | 120651.00 |
| C00341 | Geranyl diphosphate | inf | 2227.05 | 4.88 |
| C00404 | (Phosphate)n; (Phosphate)n+1; (Phosphate)n-1 | 2.91e-006 | -32.22 | 1.88 |
| C00536 | Triphosphate; Tripolyphosphate | 2.81e-006 | -32.31 | 1.85 |
| C00677 | Deoxynucleoside triphosphate | - | - | - |
| C02569 | Neryl diphosphate; Neryl pyrophosphate | 3.95e-004 | -19.81 | 3.46 |
| C03190 | (+)-Bornyl-diphosphate | 4.09e-004 | -19.72 | 4.53 |
| C03279 | Inorganic triphosphate | - | - | - |
| C04093 | poly-cis-Polyprenyl diphosphate | 5.97e-004 | -18.76 | 5.02 |
| C05308 | Linaloyl diphosphate | - | - | - |
| C05470 | Urocortisone | 2.76e-001 | -3.26 | 8.22 |
| C05806 | Polyprenyl diphosphate | 4.87e-004 | -19.28 | 5.09 |

**Table 4.8: IC50 results of ligands**

| Name | Ic50 | G bind,kj/ | Rms,A |
|---|---|---|---|
| 1.1.2130 | 3.10e+091 | 532.48 | 1.50 |
| 3.4.2130 | 1.73e+043 | 251.64 | 1.77 |
| 9.2.2130 | 8.96e+010 | 63.66 | 3.42 |
| 10.10.2130 | 4.45e+016 | 96.89 | 3.16 |
| 15.2.12645 | 1.12e+052 | 302.93 | 2.01 |
| 18.2.12645 | 4.80e+000 | 3.97 | 4.04 |
| 19.1.12645 | inf | 5619.91 | 3.21 |
| 23.6.2130 | 2.46e-002 | -9.37 | 5.28 |
| 2130.5 | 2.01e+041 | 240.38 | 2.03 |
| 2137.1 | 1.28e-005 | -28.48 | 1.09 |
| 3039.3 | 1.05e-003 | -17.34 | 1.32 |
| 2-[[4-(dimethylamino)phenyl]sulfonyl]-1-[5-nitro-2-furyl]ethanone | inf | 15862.40 | 2.99 |

**Table 4.9: Druglikeness results**

| Name | LogP | Logs | Lipinski's rule |
|---|---|---|---|
| 1.1.2130 | 4.7 | -5.4 | true |
| 3.4.2130 | 4.7 | -5.4 | true |
| 9.2.2130 | 5.1 | -5.7 | false |
| 10.10.2130 | 5.1 | -5.7 | false |
| 15.2.12645 | 0.5 | -1.7 | true |
| 18.2.12645 | -0.2 | -0.8 | true |
| 19.1.12645 | -0.2 | -1.0 | true |
| 23.6.2130 | 4.1 | -4.8 | true |
| 2130.5 | 5.4 | -5.5 | false |
| 2137.1 | 1.8 | -2.9 | false |
| 3039.3 | 1.7 | -5.2 | true |
| 12645.3 | 0.3 | -1.3 | True |

**Table 4.10: Goldscore and chemscore results**

| Name | Goldscore | Chemscore |
|---|---|---|
| 1.1.2130 | 73.46 | 8.73 |
| 3.4.2130 | 72.24 | 6.64 |
| 9.2.2130 | 71.20 | 8.79 |
| 10.10.2130 | 73.29 | 6.97 |
| 15.2.12645 | 73.16 | 27.78 |
| 18.2.12645 | 71.79 | 26.35 |
| 19.1.12645 | 71.16 | 26.83 |
| 23.6.2130 | 70.37 | 7.52 |
| 2130 | 61.77 | 7.29 |
| 2137 | 58.64 | 26.90 |
| 3039 | 61.52 | 26.96 |
| 2-[[4-(dimethylamino)phenyl]sulfonyl]-1-[5-nitro-2-furyl]ethanone | 70.34 | 29.17 |

**Table 4.11: Q-ALBUMIN RESULTS**

| Name | Albumin, pKd | Site1, pKd | Site2, pKd | LogP | Mol. weight | H-bond acceptors | H-bond Donors | Rotatable Bonds | Lipinski's rule |
|---|---|---|---|---|---|---|---|---|---|
| 1.1.2130 | N/A | N/A | N/A | 4.6 | 485.2 | 4 | 0 | 8 | Yes |
| 3.4.2130 | 3.6 | 3.4 | 3.6 | 4.7 | 483.2 | 4 | 0 | 8 | Yes |
| 9.2.2130 | 4.3 | 2 | 4.3 | 5.6 | 481.2 | 3 | 0 | 9 | No |
| 10.10.2130 | 3.9 | 1.9 | 3.9 | 5.6 | 481.2 | 3 | 0 | 9 | No |
| 15.2.12645 | 4.9 | 3 | 4.9 | 1.6 | 354.4 | 8 | 1 | 13 | Yes |
| 18.2.12645 | 5.3 | 4.5 | 5.3 | 0.8 | 355.4 | 9 | 3 | 13 | Yes |
| 19.1.12645 | 3.9 | 3.9 | 3.3 | 1 | 355.4 | 9 | 2 | 13 | Yes |
| 23.6.2130 | 4.5 | 4.5 | 3.2 | 4.8 | 455.1 | 3 | 1 | 8 | Yes |
| Kegg374 | 6 | 4.9 | 6 | -0.7 | 257.9 | 10 | 0 | 4 | Yes |
| Kegg491 | 6 | 4.9 | 6 | -0.7 | 257.9 | 10 | 0 | 4 | Yes |
| Kegg2571 | 5.3 | 4.3 | 5.3 | -0.7 | 257.9 | 10 | 0 | 4 | Yes |
| 12645.3 | 5.2 | 3.8 | 5.2 | 1.9 | 338.3 | 8 | 0 | 8 | Yes |
| 2137.1 | 3.2 | 2.3 | 3.2 | 5.4 | 439.1 | 2 | 0 | 7 | No |
| 2130.5 | 4.1 | 4.1 | 3.6 | 2.9 | 323.1 | 7 | 2 | 5 | Yes |
| 3039.3 | 4.5 | 4.1 | 4.5 | 1.1 | 391.4 | 6 | 6 | 11 | No |

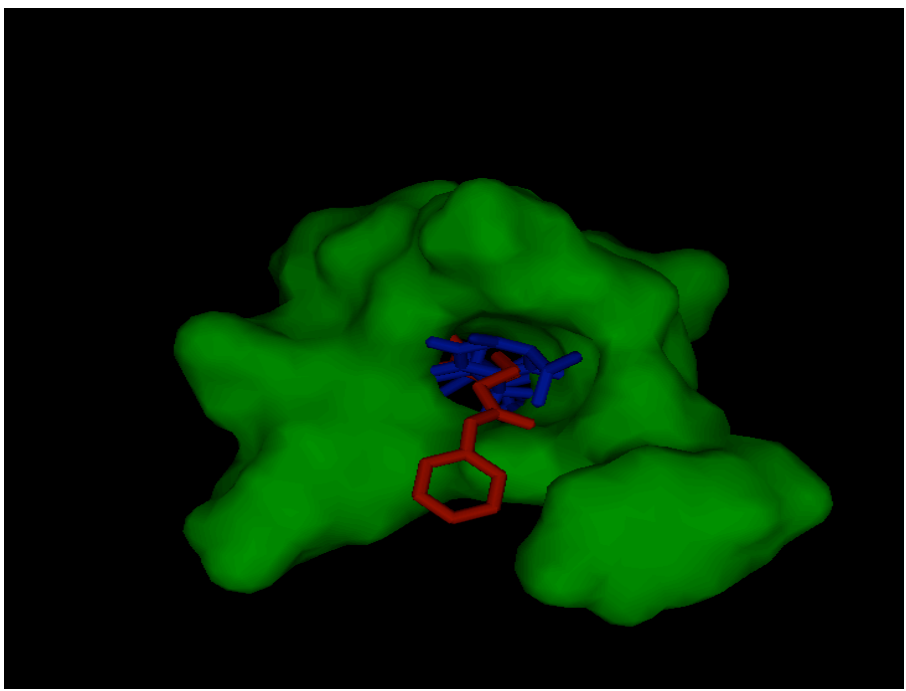

*Fig: 13: 12645 within the active site of SHH*

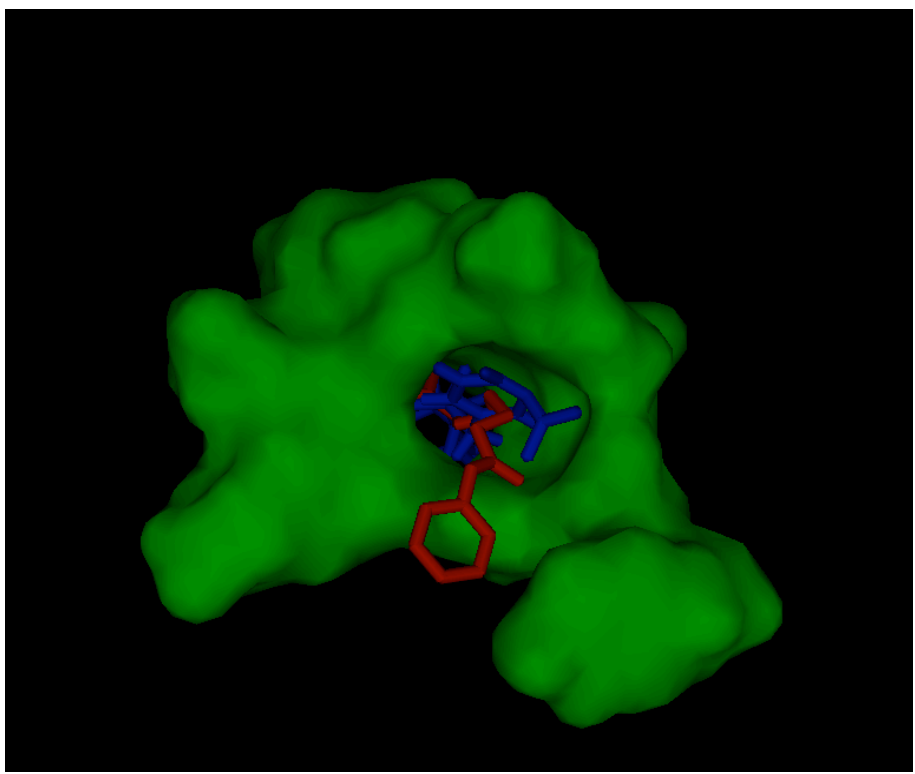

*Fig: 14: methyl group added at C-4 of 12645 within the active site of SHH*

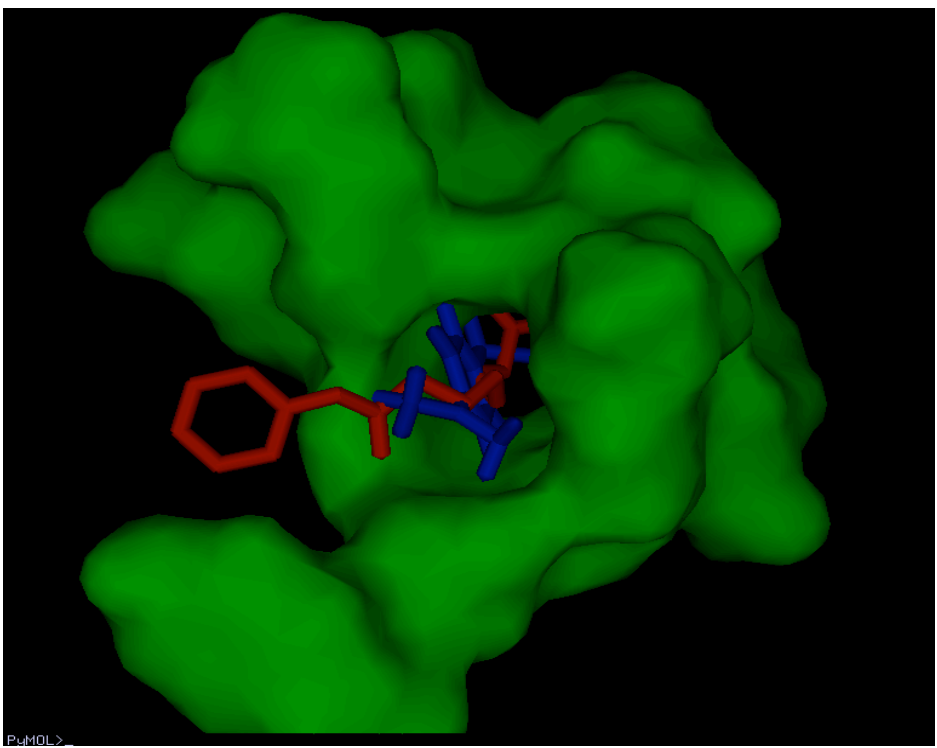

*Fig: 15: amino group added at C-5 of 12645 within the active site of SHH*

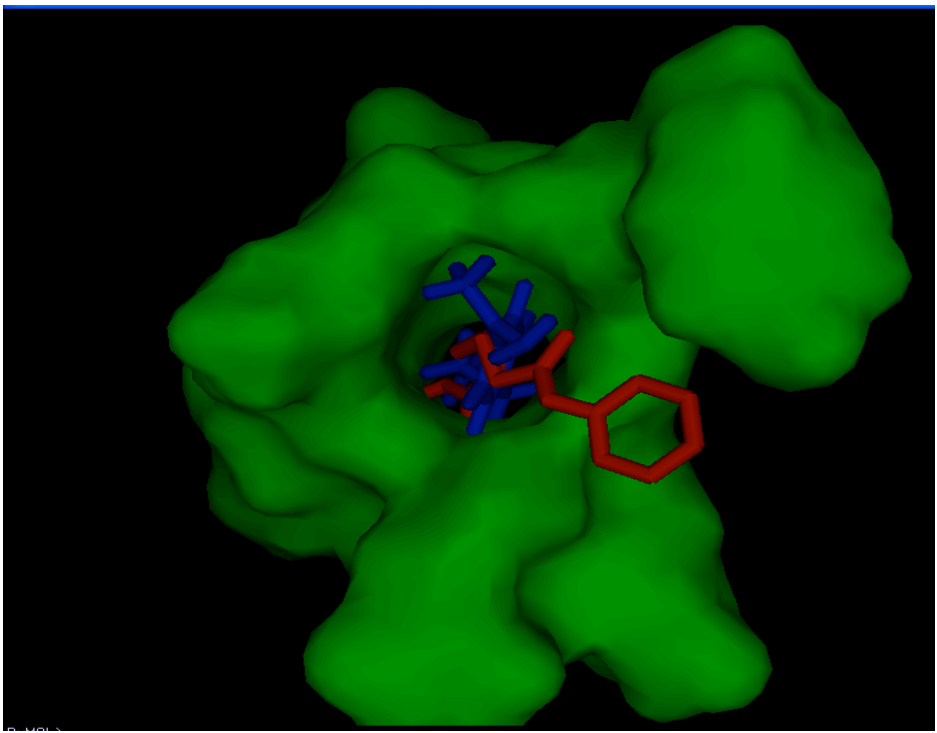

*Fig: 16: amino group added at C-4 of 12645 within the active site of SHH*

The computational experiment undertaken has resulted in the identification of a few small molecules, which docked well in to the active site of the target. Careful visual inspection of the top ranked molecules (hits) yielded a list of four small molecules. These four molecules yielded high goldscore and chemscore value, which indicates the stability of the structure. When further analyses were done using ic50, druglikeness, albumin pkd, the results were found to lie within an optimum range. The docked poses, along with their corresponding GoldScore and ChemScore are given. These molecules are suggested to be interesting candidates for further testing in the laboratory.

**Table 4.12: Novel potential ligands**

| Molecule name | Goldscore | Chemscore |
| --- | --- | --- |
| 2-[[4-(dimethylamino)phenyl]sulfonyl]-1-[5-nitro-2-furyl]ethanone | 70.34 | 29.17 |
| Methyl group added to 2-[[4-(dimethylamino)phenyl]sulfonyl]-1-[5-nitro-2-furyl]ethanone at C-4 position | 73.16 | 27.78 |
| Amino group added to 2-[[4-(dimethylamino)phenyl]sulfonyl]-1-[5-nitro-2-furyl]ethanone at C-5 position | 71.79 | 26.35 |
| Amino group added to 2-[[4-(dimethylamino)phenyl]sulfonyl]-1-[5-nitro-2-furyl]ethanone at C-4 position | 71.16 | 26.83 |

# SUMMARY AND CONCLUSION

Histone deacetylase (HDAC) and Histone acetyl-transferase (HAT) are enzymes that influence transcription by selectively deacetylating or acetylating the ε-amino groups of lysine located near the amino termini of core histone proteins. Over expression of HDACs noted in many forms of cancers including leukemia and breast cancer. There is a growing interest in the development of histone deacetylase inhibit. There is a growing interest in the development of histone deacetylase inhibitors as anti cancer agents. In this study the active site of HDAC-8 is defined as the residues which are 4 Å vicinity of the ligand. Large databases of small molecules were computationally screened using molecular docking for "hits" that can conformationally and chemically fit to the active site.

The study has identified four putative small molecular inhibitors that might bind well to the active site of the target molecule chosen for the study (HDAC-8). These molecules, predicted to "dock" well into the active site of human HDAC-8 should be considered as "interesting" molecules that need to be further tested in the laboratory. Finally, this purely insilico study strongly underscores the importance of computational approaches in drug discovery, supplementing classical methods, thus saving enormous amount of time and money.

# BIBLIOGRAPHY

1. C. M. Oshiro, I. D. Kuntz and J. Scott Dixon, Flexible ligand docking using a genetic algorithm, Journal of Computer-Aided Molecular Design, Volume 9, Number 2 April, 1995, 113-130

2. Eldridge, M.D., C.W. Murray, T.R. Auton, G.V. Paolini, and R.P. Mee. (1997). Empirical Scoring Functions. I. The Development of a Fast, Fully Empirical Scoring Function to Estimate the Binding Affinity of Ligands in Receptor Complexes. Journal of Computer-Aided Molecular Design, 11, 425-445.

3. Boehm, H-J. (1994). The Development of a Simple Empirical Scoring Function to Estimate the Binding Constant for a Protein-Ligand Complex of Known Three-Dimensional Structure. Journal of Computer-Aided Molecular Design, 8, 243-256.

4. Muegge, I. and Y.C. Martin. (1999). A General and Fast Scoring Function for Protein-Ligand Interactions: A Simplified Potential Approach. Journal of Medicinal Chemistry, 42, 791-804.

5. Gohlke, H., M. Hendlich, and G. Klebe. (2000). Knowledge Based Scoring Function to Predict Protein-Ligand Interactions. Journal of Molecular Biology, 295, 337-356.

6. Kouraklis G, Theocharis S. Histone deacetylase inhibitors: a novel target of anticancer therapy (review). 2006 Feb;15(2):489-494.

7. Fiona McLaughlin1 and Nicholas B. La Thangue, Biochemical Pharmacology, Volume 68, Issue 6, 15 September 2004, Pages 1139-1144.

8. Douglas B Kitchen et al: Docking and scoring in virtual screening for drug discovery: methods and applications, nature, 2004, vol 3.

9. A. Kim, J. H. Shin, I. H. Kim, J. H. Kim, J. S. Kim, H. G. Wu, E. K. Chie, S. W. Ha, C. I. Park, and G. D. Kao, Histone Deacetylase Inhibitor-Mediated Radiosensitization of Human Cancer Cells: Class Differences and the Potential Influence of p53, Clin. Cancer Res., February 1, 2006; 12(3): 940 – 949

10. D. Mottet, A. Bellahcene, S. Pirotte, D. Waltregny, C. Deroanne, V. Lamour, R. Lidereau, and V. Castronovo Histone Deacetylase 7 Silencing Alters Endothelial Cell Migration, a Key Step in Angiogenesis Circ. Res., December 7, 2007; 101(12): 1237 - 1246.

11. G. Garcia-Manero, H. Yang, C. Bees-Ramos, A. Ferrajoli, J. Cortes, W. G. Wierda, S. Faderl, C. Koller, G. Morris, G. Rosner, et al. Phase 1 study of the histone deacetylase inhibitor vorinostat (suberoylanilide hydroxamic acid [SAHA]) in patients with advanced leukemias and myelodysplastic syndromes Blood, February 1, 2008; 111(3): 1060 - 1066.

12. V. R. Fantin and V. M. Richon. Mechanisms of Resistance to Histone Deacetylase Inhibitors and Their Therapeutic Implications .Clin. Cancer Res., December 15, 2007; 13(24): 7237 – 7242.

13. Eyüpoglu IY, Hahnen E, Buslei R, Siebzehnrübl FA, Savaskan NE, Lüders M, Tränkle C, Wick W, Weller M, Fahlbusch R, Blümcke I. Suberoylanilide hydroxamic acid (SAHA) has potent anti-glioma properties in vitro, ex vivo and in vivo. J Neurochem. 2005 May;93(4):992-999.

14. Anthony E Dear, Hong B Liu, Penelope A Mayes, Patrick Perlmutter Australian Centre for Blood Diseases, Department of Medicine, Monash University, 6th floor Barnet Tower, 89 Commercial Rd, Prahran, 3181, Australia. Org Biomol Chem. Conformational analogues of Oxamflatin as histone deacetylase inhibitors. 2006 Oct 21;4 (20):3778-3784 17024284 (P,S,E,B,D).

15. Verdonk ML, Cole JC, Hartshorn MJ, Murray CW, Taylor RD. Improved protein-ligand docking using GOLD. Proteins. 2003 Sep 1; 52(4):609-623.

16. Matthew J. Marton1, Joseph L. DeRisi2, Holly A. Bennett1, Vishwanath R. Iyer2, Michael R. Meyer1, Christopher J. Roberts1, Roland Stoughton1, Julja Burchard1, David Slade1, Hongyue Dai1, Douglas E. Bassett Jr.1, Leland H. Hartwell3, Patrick O. Brown2 & Stephen H. Friend1. Drug target validation and identification of secondary drug target effects using DNA microarrays. Nature Medicine 4, 1293 - 1301 (1998) doi:10.1038/3282

17. Dive C, Hickman JA. Drug-target interactions: only the first step in the commitment to a programmed cell death? Br J Cancer. 1991 Jul;64(1):192-196